\documentclass[preprint,1p,times]{elsarticle}
\usepackage{algorithm}
\usepackage{algpseudocode}
\usepackage{centernot}
\usepackage{verbatim}
\usepackage{amssymb}
\usepackage{amsmath}
\usepackage{listings}
\usepackage[hidelinks]{hyperref}
\usepackage{pgfplots}
\usepackage{enumitem}
\usepackage{titlesec}
\usetikzlibrary{patterns}

\titleclass{\subsubsubsection}{straight}[\subsection]

\newcounter{subsubsubsection}[subsubsection]
\renewcommand\thesubsubsubsection{\thesubsubsection.\arabic{subsubsubsection}}

\titleformat{\subsubsubsection}
  {\normalfont\normalsize\itshape}{\thesubsubsubsection}{1em}{}
\titlespacing*{\subsubsubsection}
{0pt}{3.25ex plus 1ex minus .2ex}{1.5ex plus .2ex}

\makeatletter
\renewcommand\paragraph{\@startsection{paragraph}{5}{\z@}%
  {3.25ex \@plus1ex \@minus.2ex}%
  {-1em}%
  {\normalfont\normalsize\itshape}}
\renewcommand\subparagraph{\@startsection{subparagraph}{6}{\parindent}%
  {3.25ex \@plus1ex \@minus .2ex}%
  {-1em}%
  {\normalfont\normalsize\itshape}}
\def\toclevel@subsubsubsection{4}
\def\toclevel@paragraph{5}
\def\toclevel@paragraph{6}
\def\l@subsubsubsection{\@dottedtocline{4}{7em}{4em}}
\def\l@paragraph{\@dottedtocline{5}{10em}{5em}}
\def\l@subparagraph{\@dottedtocline{6}{14em}{6em}}
\makeatother

\pgfplotsset{compat=1.9}

\pgfplotsset{
    /pgfplots/ybar legend/.style={
    /pgfplots/legend image code/.code={%
       \draw[##1,/tikz/.cd,yshift=-0.25em]
        (0cm,0cm) rectangle (3pt,0.8em);},
   },
}

\pgfplotsset{every axis/.append style={
		font=\footnotesize,
		line width=0.6pt,
		}}

\journal{Journal of Parallel and Distributed Computing}

\begin{document}

\newenvironment{myitemize}
{ \begin{itemize} [topsep= 3pt] 
    \setlength{\itemsep}{0pt}
    \setlength{\parskip}{0pt}
    \setlength{\parsep}{0pt}     }
{ \end{itemize}                  }

\newcommand {\inred}[1] {\textcolor{red}{#1}}
\newcommand {\curle}[1]{ \left \{      #1 \right \}      }
\newcommand {\floor}[1]{ \left \lfloor #1 \right \rfloor }
\newcommand {\ceil} [1] { \left \lceil  #1 \right \rceil  }
\newcommand {\fin}  {\textbf{input:\;}}
\newcommand {\fout} {\textbf{output:\;}}
\newcommand {\nextfloat} [1] {\phi\left(#1\right)}
\newcommand {\perfchart}[2]{
	\begin{tikzpicture}
	\begin{loglogaxis}[
			height=6.5cm,
			title={#2 precision},
			xmin=8, xmax=2097152,
			ymin=3, ymax=1000,
			log basis x=2,
			log basis y=10,
			xtick={16,256,4096,65536,1048576},
			xminorticks=false,
			yminorticks=true,
			legend pos=outer north east,
			grid style=dashed,
		]
	\input{#1-#2.perfplot}
	\end{loglogaxis}
	\end{tikzpicture}
}

\newcommand \SingleName {Single}
\newcommand \DoubleName {Double}
\newcommand \ScalarName {Scalar}
\newcommand \SSEName {SSE}
\newcommand \AVXName {AVX}

\newcommand \MorinOffsetName {KhuongOffset}
\newcommand \MorinBranchyName {KhuongBranchy}
\newcommand \LowerBoundName{lower\_bound}
\newcommand \BitSetNoPadName{LeadBitNoPad}
\newcommand \ClassicModName{ClassicMod}

\newcommand \ClassicName {Classic}
\newcommand \ClassicColor {magenta}
\newcommand \ClassicMark {oplus}
\newcommand \ClassicStyle {solid}

\newcommand \ClassicOffsetName {Offset}
\newcommand \ClassicOffsetColor {red}
\newcommand \ClassicOffsetMark {*}
\newcommand \ClassicOffsetStyle {solid}

\newcommand \BitSetName {LeadBit}
\newcommand \BitSetColor {green}
\newcommand \BitSetMark {*}
\newcommand \BitSetStyle {solid}

\newcommand \EytzingerName {Eytzinger}
\newcommand \EytzingerColor {blue}
\newcommand \EytzingerMark {otimes}
\newcommand \EytzingerStyle {solid}

\newcommand \DirectName {Direct}
\newcommand \DirectColor {blue}
\newcommand \DirectMark {*}
\newcommand \DirectStyle {dotted}

\newcommand \DirectFMAName {DirectFMA}
\newcommand \DirectFMAColor {blue}
\newcommand \DirectvMark {*}
\newcommand \DirectFMAStyle {dashed}

\newcommand \DirectGapName {DirectGap2}
\newcommand \DirectGapColor {green}
\newcommand \DirectGapMark {*}
\newcommand \DirectGapStyle {dotted}

\newcommand \DirectGapFMAName {DirectGap2FMA}
\newcommand \DirectGapFMAColor {green}
\newcommand \DirectGapFMAMark {*}
\newcommand \DirectGapFMAStyle {dashed}

\newcommand \DirectCacheName {DirectCache}
\newcommand \DirectCacheColor {red}
\newcommand \DirectCacheMark {*}
\newcommand \DirectCacheStyle {dotted}

\newcommand \DirectCacheFMAName {DirectCacheFMA}
\newcommand \DirectCacheFMAColor {red}
\newcommand \DirectCacheFMAMark {*}
\newcommand \DirectCacheFMAStyle {dashed}

\newcommand \TernaryName {Ternary}
\newcommand \TernaryColor {red}
\newcommand \TernaryMark {*}
\newcommand \TernaryStyle {dashed}

\newcommand \PentaryName {Pentary}
\newcommand \PentaryColor {green}
\newcommand \PentaryMark {*}
\newcommand \PentaryStyle {dashed}

\newcommand \NonaryName {Nonary}
\newcommand \NonaryColor {blue}
\newcommand \NonaryMark {*}
\newcommand \NonaryStyle {dashed}

\newcommand \MKLName {MKL}
\newcommand \MKLColor {yellow}
\newcommand \MKLMark {diamond*}
\newcommand \MKLStyle {solid}

\begin{frontmatter}
\title{A Fast and Vectorizable Alternative to Binary Search in $O(1)$ with Wide Applicability to Arrays of Floating Point Numbers}

\author{\renewcommand*{\thefootnote}{\fnsymbol{footnote}}
	Fabio Cannizzo\footnote{DISCLAIMER: Although Fabio Cannizzo is employed by Standard Chartered at the time this paper is written, this paper has been produced by Fabio Cannizzo in a personal capacity and Standard Chartered is not associated or responsible for its content in any way.}}

\begin{abstract}
Given an array $X$ of $N+1$ strictly ordered floating point numbers\footnote{a floating point number is a rationale number which can be represented exactly on a computer given a chosen floating point representation} and a floating point number $z$ belonging to the interval $[X_0,X_N)$, a common problem in numerical methods is to find the index $i$ of the interval $[X_{i},X_{i+1})$ containing $z$, i.e. the index of the largest number in the array $X$ which is smaller or equal than $z$. This problem arises for instance in the context of spline interpolation or the computation of empirical probability distribution from empirical data. Often it needs to be solved for a large number of different values $z$ and the same array $X$, which makes it worth investing resources upfront in pre-processing the array $X$ with the goal of speeding up subsequent search operations. In some cases the values $z$ to be processed are known simultaneously in blocks of size $M$, which offers the opportunity to solve the problem vectorially, exploiting the parallel capabilities of modern CPUs.
The common solution is to sequentially invoke $M$ times the well known \textit{binary search} algorithm, which has complexity $O(log_2 N)$ per individual search and, in its classic formulation, is not vectorizable, i.e. it is not SIMD friendly. This paper describes technical improvements to the \textit{binary search} algorithm, which make it faster and vectorizable. Next it proposes a new vectorizable algorithm, based on an indexing technique,  applicable to a wide family of $X$ partitions, which solves the problem with complexity $O(1)$ per individual search at the cost of introducing an initial overhead to compute the index and requiring extra memory for its storage. Test results using streaming SIMD extensions compare the performance of the algorithm versus various benchmarks and demonstrate its effectiveness. Depending on the test case, the algorithm can produce a throughput up to two orders of magnitude larger than the classic \textit{binary search}. Applicability limitations and cache-friendliness related aspects are also discussed.
\end{abstract}


\end{frontmatter}

\section{Introduction}
\label{sec:introduction}

\subsection{Problem Description}

Given an array $X$ of strictly increasing floating point numbers $\curle{X_i}_{i=0}^{N}$  
and a floating point number $z \in [X_0,X_N)$,
a common problem in computer algorithms is to find the index of the largest number in the array $X$ which is smaller or equal than the number $z$, i.e. the index $i$ such that $z \in [X_i,X_{i+1})$.

This problem arises for instance in the context of piece-wise interpolation, where a domain $[X_0,X_N)$ 
is partitioned in sub-intervals $\curle{[X_{i},X_{i+1})}_{i=0}^{N-1}$ each associated with a different interpolation function $g_i(x)$, hence, to compute the interpolated value for a number
$z$, the index $i$ of the sub-interval containing it needs to be resolved first.
Another use case is the numerical approximation of a probability distribution from a sample population, which involves defining a partition $X$ of a domain, then counting the number of samples belonging to each interval $[X_i,X_{i+1})$.

This is a generalization of the more common problem of searching in a sorted array for a matching element and it can be used as a building block to solve it. Once the index $i$ of the sub-interval $[X_i,X_{i+1})$ containing $z$ is found, if it is known that $z \in \curle{X_i}_{i=0}^{N-1}$, then $i$ is the sought answer, otherwise it suffices to check if $z$ is equal to $X_i$. 

Often the problem needs to be solved for a large number $M$ of different values $Z_j$ given the same array $X$.
The elements of the array $Z$ may become known one at a time, requiring a separate call to the search algorithm per each element $Z_j$, or in blocks, allowing to resolve multiple indices in a single call. These two  cases are respectively referred to in the sequel as the \textit{scalar} and \textit{vectorial} problems.

\subsection{Related Work}
\label{sec:relatedwork}
 The general solution to this problem is to call $M$ times the \textit{binary search} algorithm which has complexity $O\left(log_2N\right)$. A history of the \textit{binary search} algorithm is discussed by Knuth \cite{Knuth1997}, where the first published mention in the literature is attributed to Mauchly \cite{Mauchly1946}, who in 1946 proposed a \textit{binary search} algorithm to find an exact match in a sorted array of size $N$, which has to be a power of 2. It was not until 1960 that a version of the algorithm that works for any $N$ was published by Lehmer \cite{Lehmer1960}. The next step was taken by Bottenbruch \cite{Bottenbruch1962}, who presented a variation of the algorithm that avoids a separate test for equality until the very end, thus speeding up the inner loop. Knuth \cite{Knuth1997} proposed \textit{uniform binary search}, an interesting variation of \textit{binary search} using an alternative set of state variables (see section \ref{sec:offset}).
 
Many alternatives to the \textit{binary search} algorithm have been developed to improve its performance. \textit{Fibonacci search}, proposed by Kiefer \cite{Kiefer1953} and Ferguson \cite{Ferguson1960}, has the same complexity as \textit{binary search}, but improves search time when some region of the array can be accessed faster than others. \textit{Exponential search}, proposed by Bentley et al. \cite{Bentley1976}, has complexity $O\left(log_2i\right)$ where $i$ is the sought index, but, due to increased computation cost, it becomes an improvement over \textit{binary search} only if the target value lies near the beginning of the array. 

The problem was considered a solved problem for many years, as \textit{binary search} is theoretically the algorithm with best time complexity. It is only in modern times that it regained interest because of the opportunities for technical improvements introduced by the technological advances in computer architectures, like the introduction of larger main memories, cache memories, super-scalar execution units, multi core CPUs, sophisticate branch prediction algorithms, vectorial arithmetic capabilities of CPUs (SIMD) and GPGPUs.

A variation of the algorithm, specialized to find the index of the largest numbers in the array $X$ that is smaller or equal than some number $z\in[X_0,X_N)$, is described by Press et al. \cite{NRC++} and is referred to in the sequel as the \textit{classic} version of the algorithm (see section \ref{sec:binary}). It involves unpredictable control flow branches, which incur penalties on many CPU architectures; is not vectorizable, i.e. it cannot benefit from the vectorial capabilities of modern CPUs, and is not cache memory friendly, due to irregular and unpredictable data accesses. Modern research has been focusing on re-organizing code to eliminate code flow branches, exploiting parallelism and improving memory access patterns.

Zhou et al. \cite{Zhou2002} proposed a version of the algorithm which uses SIMD instructions to solve the scalar problem. The time complexity of their algorithm is $O(log_2N-log_2d)$, where $d$ is the number of floating point numbers which can be processed simultaneously\footnote{$d$ depends on the chosen set of vectorial instructions and floating point representation (e.g. with SSE instructions in \textit{single} precision\footnote{for a description of \textit{single} and \textit{double} precision floating point representation see \cite{IEEE754}} $d=4$)}, hence the improvement is noticeable only with arrays $X$ of small size.

Sanders et al. \cite{Sanders2004} described a sorting algorithm \textit{Super Scalar Sample Sort}
(\textit{SSSS}) that requires a subroutine for locating values in a sorted array, which they efficiently reorder with a special layout due to \textit{Eytzinger} \cite{Eytzinger1590}, and show that branch-free code can be used for finding exact matches.

Schlegel et al. \cite{kary2009} introduced a \textit{k-ary search} algorithm, which at every iteration subdivides the array $X$ in $k$ parts and performs $k-1$ comparisons, thus solving the problem in $log_kN$ iterations. The total number of comparisons is $(k-1)\,log_kN$, larger than $log_2N$ required by \textit{binary search}. To neutralize this extra cost they proposed to use SIMD instructions to solve the scalar problem and set $k$ to $d+1$, thus performing the $d$ comparison in parallel. To avoid the need to gather data from scattered memory locations, they re-arranged the array $X$ as in a breadth first linearized k-ary tree having $k-1$ elements in each node. This is a generalization of the \textit{Eytzinger layout} which is a special case of this layout with $k=2$. Since the algorithm requires the array $X$ to have perfect size $N=k^h-1$ for some integer $h>0$, for other values of $N$ they proposed a technique to define a conceptual \textit{complete} tree. Zeuch et al. \cite{kary2014} revisited this approach and proposed instead to pad the array at the end with its last element $X_N$, which simplifies the logic of the inner loop. This padding technique is also adopted in other algorithms discussed in this paper.

Kaldewey et al. \cite{Kaldewey2009} presented \textit{p-ary search}, which exploits gather instructions on GPGPUs to solve the vectorial problem achieving superior throughput to \textit{binary search} despite having longer latency per search.

Pulver \cite{Pulver2011} proposed an improved version of \textit{binary search}, which involves only essential arithmetic operations, is almost branch free and performs optimally when the array $X$ has size $N=2^h-1$ for some integer $h>0$.

Since for large $N$ the \textit{binary search} algorithm is memory access bound, not computation bound, significant efforts have gone into studying data layouts that could improve data locality, for example by Niewiadomski et al. \cite{Niewiadomski2006}, Bender et al. \cite{Bender2005}, Graefe et al. \cite{Graefe2001}, Rao et al. \cite{Rao1999}. 

Kim at al. \cite{Kim2011} introduced FAST, a binary tree optimized for architectural features like page size, cache line size, and SIMD bandwidth of the underlying hardware to improve memory access patterns. They used SSE instructions to solve the scalar problem performing 3 comparisons in parallel assuming an array $X$ with elements of 32 bits and processing in parallel two levels of the tree in each iteration. The weaknesses of this approach are that the bandwidth of SIMD registers is not optimally utilized and that it poses restrictions on the data type length. To obviate, they proposed an order-preserving compression technique to deal with longer data types, but apart from introducing extra computation cost, this is only applicable to an exact match search and anyway not suitable for floating point numbers.

Recently, Khuong et al. \cite{Morin2015} proposed a branch free variation of Knuth's \textit{uniform binary search} and compared its performance with existing algorithms based on alternatives layouts of the array $X$. They explored the \textit{Eytzinger layout} (Sanders et al. \cite{Sanders2004}), the \textit{(B+1)-ary trees layout} (Jones et al. \cite{Jones1986} and La Marca et al. \cite{LaMarca1996}) and the \textit{Emde Boas layout} (Frigo et al. \cite{Prokop1999}) and concluded that for small $N$ a good implementation of \textit{binary search} is faster, while for large $N$ the \textit{Eytzinger layout} performs better. Their observations contradict those previously made by Brodal et al. \cite{Brodal2002}, who conducted similar experiments and concluded that the \textit{Emde Boas layout} is faster. They also reported as beneficial the use of explicit memory pre-fetching, instructing the CPU at every iteration to speculatively load in cache the two possible memory locations that could be accessed next.

Hash based algorithms (see Ross \cite{Ross2007}) are a valid alternative when searching for exact matches, but, since they re-arrange the $X$ values in randomized order, they are not suitable for the problem described in this paper. Furthermore, robust hashing of floating point numbers, where different binary representations can refer to the same number, is not a trivial task.

In some special cases, when either the $X_i$ or the $Z_j$ numbers exhibit particular patterns, more efficient algorithms are available. 
Examples are when the numbers $X_i$ are equally spaced or when the numbers $Z_j$ are sorted, 
and the problem can be solved with complexity $O(M)$ and $O(M+N)$ respectively. However no generic alternative exists.

Functions which solve variations of the problem discussed in this paper are available in many software libraries, for instance the \textit{Intel Math Kernel Library}, the \textit{Numerical Algorithms Group Library} and the \textit{C++ Standard Library}.


\subsection{Contribution}
This paper describes technical improvements to two known variations of the \textit{binary search} algorithm, which makes them faster and vectorizable. The complexity of the algorithms remains $O\left(log_2N\right)$ per search, but performance improves by a proportionality factor $\alpha<1$. SIMD instructions are used to solve the vectorial problem, which is an approach that was not considered before, except for GPGPUs, because of the unavailability of efficient techniques to gather data from different memory locations (see Kim et al. \cite{Kim2011}).

Next it proposes a new vectorizable algorithm based on an indexing technique, which solves the problem with time complexity $O(1)$ and requires only two memory accesses, at the cost of introducing an initial overhead to compute the index and requiring extra memory for its storage. Although the algorithm has fairly wide applicability, there are particular situations where the layout of the array $X$ might cause the construction of the index to fail, thus making the algorithm inapplicable. These limitations are analyzed and discussed and variations of the algorithm which mitigate them at the cost of sacrificing some performance are introduced.

Two more variations of the algorithm are also proposed: a cache-friendly version, which improves performance for large arrays $N$ storing all the data necessary to resolve a query in a single cache line, and one which exploits modern \textit{fused multiply add} (FMA) instructions.

\section{Formal Problem Statement and Assumptions}
\label{sec:definition}
\noindent Let: 
\begin{itemize}
	\item $X$ be an array of $N+1$ floating point numbers sorted in ascending order, i.e. $X_{i}<X_{i+1}, \; i=0 \dots N-1$.
	The array $X$ is assumed to be stored in a container supporting access by index in constant time. The numbers in $X$ exhibit no special pattern.
	\item $Z$ be an unordered sequence of $V$ floating point numbers in the domain $\left[X_0, X_N \right)$.
\item $I$ be an array of $V$ unknown indices in $[0,N-1]$ such that $Z_j \in [X_{I_j}, X_{I_j+1})$
\end{itemize}
the problem consists in finding the unknown indices $I$.

It is assumed that $V$ is large and memory is not scarce, therefore it is worth investing up-front in a preliminary analysis of the structure of the array $X$ and in the creation of an auxiliary data structure in order to later achieve superior computational performance when searching for the indices $I$.

Two variations of the problem are considered:
\begin{itemize}
\item \textit{scalar search problem}: the elements of the array $Z$ become known one at a time, requiring a separate call to the search algorithm per every element $Z_j$
\item \textit{vectorial search problem}: the elements of the array $Z$ become known in blocks of size $M$, therefore allowing a block of values $Z_j$ to be passed to the search algorithm in each call.
\end{itemize}
This classification is not to be confused with the concept of \textit{scalar} versus \textit{vectorial} implementations, which refers to the utilization of vectorial capabilities of modern CPUs. Although in the sequel a \textit{vectorial implementation} is often used to solve the \textit{vectorial search problem} and a \textit{non-vectorial implementation} is used to solve the \textit{scalar search problem}, this is not always the case.

\section{Binary Search}

\subsection{\ClassicName\ Binary Search}
\label{sec:binary}

The classic implementation of \textit{binary search} described in algorithm \ref{alg:naivealg} is the one proposed in Press et al. \cite{NRC++}. It has the following weaknesses: 

\begin{algorithm}
\caption{{\ClassicName}  Binary Search (scalar problem)}
\label{alg:naivealg}
\begin{algorithmic}
\Function {\ClassicName}{\fin $z$,  $\curle{X_i}_{0}^{N}$, \fout $i$}
\State $low\; \leftarrow 0$
\State $high  \leftarrow N$
\While {$high-low > 1$} \Comment {terminating condition depends indirectly on $z$}
    \State $mid \leftarrow (low+high) / 2$
    \If {$z < X_{mid}$}  \Comment {code branch}
        \State $high \leftarrow mid$
    \Else
        \State $low \leftarrow mid$
    \EndIf
\EndWhile
\State $i \leftarrow low$
\EndFunction
\end{algorithmic}
\end{algorithm}

\begin{myitemize}
	\item the body of the loop contains a branch and each of the two possible code paths have equal probability\footnote{assuming $z$ has equal probability to be in any of the sub-intervals $\curle{[X{i},X_{i+1})}_{i=0}^{N-1}$}, which makes branch prediction algorithms used in modern CPUs ineffective; 
	\item the algorithm is not easily vectorizable because the boolean condition $z<X_{mid}$ may evaluate differently for different values of $z$ causing the program flow to take different code paths and possibly requiring a different number of iterations for the loop to complete;
	\item such uncertainty in the number of iterations is also detrimental to the efficiency of branch prediction algorithms;
    \item the loop terminating condition depends on the state variables of the loop updated after the boolean condition has been resolved, which introduces a sequentiality constraint and disables optimizations associated with out of order execution in super scalar CPUs;
    \item the memory access pattern is irregular and unpredictable, which for large $N$ causes many cache misses.
\end{myitemize}
A more detailed discussion on these issues is available in Kim et al. \cite{Kim2011}, Pulver \cite{Pulver2011}, Khuong \cite{Khuong2012} and Khuong et al. \cite{Morin2015}.

Algorithm \ref{alg:naivealg} can be made branch free and vectorizable by replacing the control flow branch in the loop body with two conditional assignments and fixing the number of iterations to the worst possible case $\ceil{log_2N}$, because performing extra unnecessary loop iterations does not alter the result. 
This variation of the algorithm, referred to in the sequel as \textit{\ClassicModName}, is a significant improvement on algorithm \ref{alg:naivealg}, because conditional assignments, although more expensive than regular assignments, are much cheaper than branch mis-predictions.

\subsection{Leading Bit Binary Search}
\label{sec:optimbinary}

Pulver \cite{Pulver2011} proposed an improved version of \textit{binary search} which requires only one conditional assignment and simplifies the arithmetic in the loop.

\begin{algorithm}[ht]
	\caption{Pulver's Leading Bit Binary Search (scalar problem)}
	\label{alg:revisited}
	\begin{algorithmic}
		\Function {PulverLeadBit}{\fin $z$,  $\curle{X_i}_{i=0}^{N}$, $P$, \fout $i$} \Comment{$P=2^{\floor{ log_2 N }}$}
		\State $i \leftarrow 0$
		\State $k \leftarrow P$
		\Repeat
		\State $r \leftarrow i\;|\;k$ \Comment{bitwise OR}
		\If {$r < N\; \&\& \; z \geq X_{r}$} \Comment{short boolean evaluation}
		\State $i \leftarrow r$
		\EndIf
		\State $k \leftarrow k / 2$ \Comment{bitwise right shift}
		\Until{ $k=0$ }
		\EndFunction
	\end{algorithmic}
\end{algorithm}

Let $p$ be the number of bits necessary to represent the number $N$, which is $p=1+\floor{ log_2 N }$,
    $b_k$ be the binary value taken by the $k$-th bit of the sought index $i$,
    $c_k = 2^{k-1}$
    and $a_k = b_kc_k$,
the index $i$ has binary representation $\sum_{k=1}^{p}a_k$.
The bits of the index can be resolved one by one starting from the highest order one as follows:
first the index is set to zero, then if $z \geq X_{c_p}$ its $p$-th bit is set, i.e. $b_p=1$,
next, if $z \geq X_{a_p+c_{p-1}}$ then its $(p-1)$-th bit is set, i.e. $b_{p-1}=1$, 
and so on, the values of the remaining bits are obtained iterating the procedure.
This methodology is described in algorithm \ref{alg:revisited}, where the input argument $P$ is the precomputed constant $$P=c_p=2^{\floor{ log_2 N }}$$
The number of iterations is fixed to $p$, therefore no longer dependent on $z$. However,
as the algorithm proceeds the candidate index of the vector $X$ could exceed the
size of the vector, therefore a double boolean condition with short circuit evaluation is necessary. If the size of the array $X$ is exactly $2^P-1$, the branch prediction algorithm learns very quickly to guess that this condition is false and the algorithm performs very well. However, the mere existence of a code flow branch makes it not vectorizable.

\subsubsection{Vectorizable Leading Bit Binary Search}
\label{sec:leadbit}
The first boolean condition in algorithm \ref{alg:revisited} can be avoided with a simple trick.
Noting that $z<X_N$ by assumption, if the array $X$ is extended to the right side padding 
it with the last entry $X_N$ up to a size which includes the largest possible index representable with $p$ bits, i.e. $2^p$, the condition $z \geq X_{r}$ resolves to \textit{false} for any $r>N$ generated by the algorithm and the corresponding bit of the index is not set.

Furthermore, the fastest scenario in which the index $r$ can grow is when the condition $z \geq X_{r}$ is false at every iteration and at the $u$-th iteration the index tested has value $r=\sum_{k=1}^u P/2^{k-1}$. When this becomes larger or equal to $N$, the condition $X_r \geq X_N$ becomes false, the corresponding bit is not set and any subsequent index generated is smaller, implying that this is the largest possible index which can ever be queried by the algorithm. 
This means that it is sufficient to extend the array $X$ only up to size $Q$
$$Q=1+\sum_{k=1}^{\bar{u}} P/2^{k-1}\quad\text{where}\quad\bar{u}=Inf \curle{u: \sum_{k=1}^u P/2^{k-1} \geq N}$$

\begin{algorithm}[ht]
	\caption{\BitSetName\  Binary Search (scalar problem)}
	\label{alg:binaryopt}
	\begin{algorithmic}
		\Function {\BitSetName}{\fin $z$,  $\curle{X_i}_{i=0}^{Q}$, $P$, \fout $i$} \Comment{$P=2^{\floor{ log_2 N }}$}
		\State $i \leftarrow 0$
		\State $k \leftarrow P$
		\Repeat
		\State $r \leftarrow i\;|\;k$ \Comment{bitwise OR}
		\If {$z \geq X_{r}$} 
		\State $i \leftarrow r$    \Comment{conditional assignment}
		\EndIf
		\State $k \leftarrow k / 2$ \Comment{bitwise right shift}
		\Until{ $k=0$ }
		\EndFunction
	\end{algorithmic}
\end{algorithm}

This yields algorithm \ref{alg:binaryopt}, which still has complexity $O(log_2N)$, but is vectorizable because the number of iterations is fixed and it involves only one conditional assignment which can be implemented without code flow branches on modern CPUs. In a vectorial implementation SIMD comparison instructions set the result register to either an \textit{all-zeros} or a \textit{all-one} bit mask and this particular type of conditional assignment can be implemented even more efficiently with just a bitwise \textit{OR} and a bitwise \textit{AND}, i.e. $i \leftarrow i \; OR \; (r \; AND \; (z \geq Z_r))$.

The associated extra memory requirement is less than or equal to $(2^P-N)\,S$ bytes, where $S$ is the size in bytes of the elements of array $X$, i.e. 4 for single precision and 8 for double precision.

\subsubsection{Vectorizable Leading Bit Binary Search With No Padding}
If memory is scarce, an alternative to extending and padding the array $X$ as in algorithm \ref{alg:binaryopt} is to take the minimum at every iteration between $r$ and $N$ as the index of the element in $X$ used for comparison.
Although the minimum operator can be resolved by the compiler without branching as a conditional assignment, it still leads to a substantial loss of performance. The extra check however is redundant and can be eliminated in the first few iterations, because the fastest possible way in which $r$ can grow is when the condition $z \geq X_r$ is always true and at the $u$-th iteration $r$ takes the value $\sum_{k=1}^u P/ 2^{k-1}$. Therefore, let $$U= Sup \curle{ u: \sum_{k=1}^u P/2^{k-1} \leq N} $$ it is guaranteed that in the first $U$ iterations $r \leq N$.

The detailed search procedure is listed in algorithm \ref{alg:binaryoptnopad}.

\begin{algorithm}[ht]
\caption{Leading Bit Vectorizable Binary Search With No Padding (scalar problem)}
\label{alg:binaryoptnopad}
\begin{algorithmic}
\Function {\BitSetNoPadName}{\fin $z$,  $\curle{X}_{i=0}^{N}$, $P$, $U$, \fout $i$}
\State $i \leftarrow 0$
\State $k \leftarrow P$

\Repeat
       \State $r \leftarrow i\;|\;k$ \Comment{bitwise OR}
	\If {$z \geq X_{r}$}   \Comment{peel first iteration}
        	\State $i \leftarrow r$
	\EndIf
	\State $k \leftarrow k/2$  \Comment{bitwise right shift}
   \State $U \leftarrow U-1$
\Until{$U > 0$}

\While{$k > 0$}
    	\State $r \leftarrow i\;|\;k$ \Comment{bitwise OR}
    	\State $w \leftarrow \min(r,N)$  \Comment{conditional assignment}
    	\If {$z \geq X_{w}$}
        	\State $i \leftarrow r$    \Comment{conditional assignment}
    	\EndIf
	\State $k \leftarrow k / 2$ \Comment{bitwise right shift}
\EndWhile
\EndFunction
\end{algorithmic}
\end{algorithm}

\subsection{Offset-Based Binary Search}
\label{sec:offset}
Khuong et al. \cite{Morin2015} proposed a quite effective variation of the classic \textit{binary search} algorithm, where the loop state variables are the \textit{base pointer} (i.e. a pointer to the first element of the sought range) and the \textit{range size}, as opposed to the \textit{start index} and the \textit{end index} used in algorithm \ref{alg:naivealg}. This allows performing only one conditional assignment to update the \textit{base pointer} at each iteration, while the \textit{range size} decreases by a constant factor $1/2$. The algorithm is branch-free and the number of iterations is fixed, which improves the effectiveness of the branch predictor.

This is a variation of Knuth's \textit{Algorithm U} (see \cite{Knuth1997}), which uses as first coordinate the \textit{start index} instead of the \textit{base pointer} and allows for the possibility of an early termination.  


Algorithm \ref{alg:naiveoffset} presented here is similar to Khuong's one, with the difference being that it uses the original state variables proposed in Knuth's \textit{Algorithm U}. This allows more flexibility when vectorizing, because the size of pointers is determined by the memory model in use, whereas the size of array indices can be tuned to fit the floating point data size, as long as the size of the array $X$ allows it. For example, working with SIMD instructions in single precision on a 64-bits platform, since floating point numbers have size 32 bits, it is more efficient if the array indices also have size 32 bits, whereas pointers have size 64 bits.


The search function requires in input the following pre-computed constants:
$$
	\begin{array}{lll}
		F=\floor{(N+1)/2} & & \text{mid index}  \\
		S=N+1-F  & & \text{mid size} \\
		J=\floor{\log_2(N+1)} & & \text{number of iterations}
	\end{array}
$$

\begin{algorithm}[ht]
	\caption{Offset Based Binary Search  (scalar problem)}
	\label{alg:naiveoffset}
	\begin{algorithmic}
		\Function {\ClassicOffsetName}{\fin $z$,  $\curle{X_i}_{i=0}^{N}$, $F$, $S$, $J$ \fout $i$}
		
		\State $i \leftarrow 0$
		\If {$z \geq X_F$ }  \Comment{assumes at least one iteration, i.e. $J>0$}
			\State $i \leftarrow F$ \Comment{conditional assignment}
		\EndIf
		\While {$J>0$}
			\State {$J \leftarrow J-1$}
			\State {$H \leftarrow \floor{S/2}$} \Comment{bitwise shift}
			\State {$F \leftarrow i+H$}
			\If {$z \geq X_F$}  \Comment{conditional assignment}
				\State {$i \leftarrow F$}
			\EndIf
			\State {$S \leftarrow F-H$}
		\EndWhile
		\EndFunction
	\end{algorithmic}
\end{algorithm}

\subsection{Cache Friendly Binary Search}
\label{sec:eytzinger}
The degree of efficiency in the use of cache memory for various formulation of comparison based search algorithms has been studied systematically in Khuong et al. \cite{Morin2015}. Their conclusion is that the most cache friendly implementation of the the binary search algorithm is the one proposed by Sanders et al. \cite{Sanders2004}, which reorders the array $X$ according with a special layout due to \textit{Eytzinger} \cite{Eytzinger1590}.

The algorithm incurs some upfront setup cost to rearrange the array $X$ and requires extra memory storage space to store it once reordered. An efficient implementation requires that the size of the original array $X$ is $2^h-1$ for some integer value $h>0$. 

Algorithm \ref{alg:eytzinger} describes the index search procedure, assuming that the Eytzinger layout has already been computed and stored in a new array $Y$. 
While Sanders' algorithm, as implemented in Khuong at al. \cite{Morin2015}, can only cope with arrays of size $2^h-1$, here the array $X$ is padded with the last element of the array $X_N$ until it reaches the size $2^h-1$, so that the algorithm does not require the introduction of checks for the index range and can work at its best efficiency. This trick, already used in algorithm \ref{alg:binaryopt}, does not affect results, because by the problem statement, the condition $z<X_N$ is always true.

This introduces an extra memory requirement of about $(2^L-N)\,S$ bytes, where $S$ is the size in bytes of the elements of array $X$, i.e. 4 for single precision and 8 for double precision. Noting that some nodes in the tree may never be reached, the size of the padded array $Y$ could be reduced, similarly to what was done in section \ref{sec:leadbit}.

\begin{algorithm}[ht]
	\caption{Eytzinger Binary Search (scalar problem)}
	\label{alg:eytzinger}
	\begin{algorithmic}
		\Function {\EytzingerName}{\fin $z$, $\curle{Y_i}_{i=0}^{2^{L+1}-2}$, $M$, $L$, \fout $i$} 
		\State {$P \leftarrow 1$}
		\If {$z \geq Y_0$}
		\State {$P \leftarrow 2$}  \Comment {conditional assignment}
		\EndIf
		\While {$L>1$}
		\State {$Q \leftarrow 1$}
		\If {$z \geq Y_P$}
		\State {$Q \leftarrow 2$}  \Comment {conditional assignment}
		\EndIf
		\State {$P \leftarrow 2P + Q$}
		\State {$L \leftarrow L-1$}
		\EndWhile
		
		\State {$i \leftarrow P \, \& \, M $} \Comment {bitwise AND}
		
		\EndFunction
	\end{algorithmic}
\end{algorithm}

The detailed search procedure is listed in algorithm \ref{alg:eytzinger}, which uses the following precomputed constants:
\begin{align*}
	L&=1+\floor{\log_2(2+N)} \\
	M&= not \; (2L) \quad (not \text{ is the bitwise not operator})
\end{align*}

In a vectorial implementation with SIMD instructions, since comparison operations set the result register to either an \textit{all-zeros} or an \textit{all-ones} bit mask, which correspond respectively to the integer numbers $0$ and $-1$, the two conditional assignments in algorithm \ref{alg:eytzinger} can be rewritten more efficiently with just an integer subtraction, e.g. $Q \leftarrow Q - (z \geq Y_P)$.


\section{Direct Search}
\label{sec:directmethod}
A scalar algorithm with complexity $O(1)$ per individual search can be obtained via construction of an auxiliary function which maps real numbers $z\in[X_{0},X_{N})$ directly to the sought indices $i$ according with the following simple procedure.

Let $f$ be a function which maps floating point numbers $z\in[X_{0},X_{N}]$ to natural numbers in $[0,R]$ for some value $R \geq N$ and satisfies the following properties
\begin{subequations}
\label{eq:fprop}
\begin{align}
&f(X_0)=0 \\
&f(X_N)=R \label{eq:lastf} \\
&\forall a,b \in[X_{0},X_{N}], \; a>b \implies f(a) \geq f(b) \label{eq:monotonic} \\
&f(X_{i+1}) > f(X_i), &&\quad  i=0 \dots N-1 \label{eq:f-monotonic}
\end{align}
\end{subequations}
For example, if the attention was restricted to arrays $X$ such that $\curle{X_{i+1}-X_i}_{i=0}^{N-1} \geq 1$, a function $f$ satisfying these properties could be the \textit{floor} operator $f(z)=\floor z$. \\
Let $K$ be a sorted array of size $R+1$ of natural numbers mapping the indices $j$ generated by function $f$ to indices of the array $X$ as follows
\begin{equation}
\label{eq:indexdef}
K_{j}=\left\{ 
	\begin{array}{ll}
		0, & j=0 \\
		i, & f(X_{i-1}) < j \leq f(X_i)
	\end{array}, \quad j=0\dots R
\right.
\end{equation}
Possible pseudo-code to construct the array $K$ as specified in \eqref{eq:indexdef} is proposed in algorithm \eqref{alg:initk}.
\begin{algorithm}[ht]
	\caption{Initialization of array $K$ (pseudo-code)}
	\label{alg:initk}
	\begin{algorithmic}
		\Function {InitK}{\fin $\curle{X_i}_{i=0}^{N}$, $f(z)$, \fout $\curle{K_j}_{j=0}^{R}$}
		\State {$b \leftarrow R$}
		\State {$i \leftarrow N$}
		\Repeat
			\State {$t \leftarrow f(X_i)$}
			\While {$b > t$}  \Comment{always \textit{false} at the first iteration, when $i=N$ and $b=R$}
				\State {$K_b \leftarrow j$}
				\State {$b \leftarrow b-1$}
			\EndWhile

			\State {$j \leftarrow i$}   		\Comment{at the first iteration, when $i=N$, $j$ is initialized here}
			\State {$K_b \leftarrow j$}
			\State {$b \leftarrow b-1$}
		
			\State $i \leftarrow i-1$
		\Until{ $b < 0$ }
		\EndFunction
	\end{algorithmic}
\end{algorithm}

\noindent The definition \eqref{eq:indexdef} of array $K$, implies the following property
\begin{align}
\label{eq:kidentity}
   c=f(X_i) \; \implies \; K_c=i
\end{align}
Given a floating point number $z\in[X_{i},X_{i+1}\,)$, property \eqref{eq:monotonic} of function $f$ guarantees that
$$
 	a=f(X_{i}) \leq j=f(z) \leq f(X_{i+1})=b,
$$
and property \eqref{eq:f-monotonic} implies that $a<b$. Because of \eqref{eq:kidentity} $K_a=i$ and $K_b=i+1$, hence the index $t=K_j$ can be either $i$ or $i+1$
$$i = K_a \leq t=K_j \leq K_b = i+1$$
Since $z\in[X_{i},X_{i+1}\,)$, if $t=i$ then $z \geq X_t$, while if $t=i+1$ then $z<X_t$, therefore the sought index can be trivially resolved comparing $z$ with $X_t$:
$$
i = 
\left\{
	\begin{array}{ll}
		t-1 & \text{if } z < X_t,  \\
		t   & \text{otherwise}
	\end{array}
\right.
$$

\subsection{Proposed Choice for the Function $f$}
\label{sec:constructf}
The search procedure described in section \ref{sec:directmethod} relies on the existence of a function $f$ satisfying properties \eqref{eq:fprop}. Since the largest possible value generated by the function is $R$, which defines the size of the array $K$, in order to minimize storage space requirements and initialization cost, it is desirable for $R$ to be as small as possible. Furthermore, since function $f$ is used in the search routine, it is desirable for it to be computationally fast. \\

A possible choice, not necessarily optimal, is to use the simple formula
\begin{align}
\label{eq:function}
	f(z) = \floor{ H\, (z-X_0) }
\end{align}
where $H$ is an appropriately chosen constant.  This is clearly a computationally efficient function, as its evaluation requires only a multiplication, a subtraction and a truncation.

Condition \eqref{eq:f-monotonic} requires that $H$ satisfies the inequalities
\begin{align}
\label{eq:fcond}
	\floor{H\,(X_{i+1}-X_0)} > \floor{H\,(X_i-X_0)}, \quad  i=0 \dots N-1
\end{align}
and the truncation operation can be removed writing the more restrictive system of inequalities
\begin{align}
\label{eq:fcond9}
	H\,(X_{i+1}-X_0) > 1+ H\,(X_i-X_0), \quad  i=0 \dots N-1
\end{align}
which yields the theoretical lower bound
\begin{equation}
\label{eq:minh}
	H > \frac{1}{\min_i\curle{X_{i+1}-X_i}_{i=0}^{N-1}}
\end{equation}
Once a value for $H$ satisfying the lower bound \eqref{eq:minh} is chosen, $R$ is simply determined applying formula \eqref{eq:function} to $X_N$
\begin{equation}
\label{eq:minr}
	R = \floor{H\,(X_{N}-X_0)}
\end{equation}
A possible value for the constant $H$ which strictly satisfies inequality \eqref{eq:minh} could be obtained as
\begin{subequations}
\label{eq:rhtheory}
\begin{align}
\label{eq:rtheory}
	R &= 1+\ceil { \frac{X_N-X_0}{\min_i\curle{ X_{i}-X_{i-1} }_{i=1}^{N} } } \\
\label{eq:htheory}
        H &= \frac{R}{X_N-X_0}
\end{align}
\end{subequations}
In reality, as explained in section \ref{sec:rounding}, because of rounding errors this is not a robust approach to determine $H$ and a different methodology must be used.

Note that, because of the transformations performed on inequalities \eqref{eq:fcond} into more restrictive conditions, the lower bound $H$ proposed in \eqref{eq:minh} is not guaranteed to be optimal, i.e. it is not guaranteed to be the smallest possible value which would make function \eqref{eq:function} compatible with properties \eqref{eq:fprop}. For instance, given the array $X=\{0, 0.5, 0.7, 1.1\}$, equations \eqref{eq:minh} yields $H \approx 6.36$, however smaller values would also be admissible, e.g. $H=3$. The determination of the optimal value of $H$ satisfying properties \eqref{eq:fprop} is not addressed in this paper.

\subsection{Algorithm}
\label{sec:dirsearch}

Summarizing, using the choice for function $f$ proposed in \eqref{eq:function}, for a given number $z\in [X_0,X_N)$ the index $i$ such that $z\in [X_i,X_{i+1})$ can be obtained with the following procedure, given in pseudo-code in algorithm \ref{alg:direct}:
\begin{enumerate}
	\item compute the index $j$ using \eqref{eq:function}
	\item read the correspondent index $i$ stored in $K_j$
	\item if the number $z$ is smaller than $X_i$, then decrease the index $i$ by one.
\end{enumerate}
Note that the last step does not involve any conditional jump. In the scalar case it can be resolved via conditional assignment. In the vectorial case with streaming SIMD extensions a floating point comparison operation returns a bit mask, which, if reinterpreted as a signed integer, is either $0$ or $-1$ and can be trivially added to $i$.

\begin{algorithm}
	\caption{Direct Search (scalar problem)}
	\label{alg:direct}
	\begin{algorithmic}
		\Function {\DirectName}{\fin $z$, $\curle{X_i}_{i=0}^{N}$, $\curle{K_j}_{j=0}^{R}$, $H$ \fout $i$}
		\State $j \leftarrow \floor{ H \, (z-X_0) }$
		\State $i \leftarrow K_j$
		\If {$z<X_{i}$}
			\State $i \leftarrow i-1$ \Comment {conditional assignment}
		\EndIf
		\EndFunction
	\end{algorithmic}
\end{algorithm}

Similarly to what was already performed in section \ref{sec:eytzinger}, working with SIMD instructions, the conditional assignment can be implemented with just with just an integer addition, e.g. $i \leftarrow i + (z \geq X_i)$.

\subsection{Geometric Interpretation}
\label{sec:geometric}

The algorithm described in section \eqref{sec:dirsearch} has an intuitive equivalent geometric interpretation.
Let $\curle{Y_j=X_0+j/H}_{j=0}^{R+1}$ be a conceptual array of equally spaced real numbers, the segments $\curle{[Y_{j},Y_{j+1})}_{j=0}^{R}$ have constant length $1/H$ and overlap with the segments $\curle{[X_{i},X_{i+1})}_{i=0}^{N-1}$.

\noindent The function $j=f(z)$ defined in \eqref{eq:function} can be interpreted as the mapping from a real number $z\in [X_0,X_N)$ to the interval $[Y_{j},Y_{j+1})$ which contains it. 

\noindent The array $K$ defined in \eqref{eq:indexdef} can be interpreted as the mapping of the abstract segments $\curle{[Y_{j},Y_{j+1})}_{j=0}^{R}$ to elements of the array $X$ as re-defined below and illustrated in figure \ref{pic:directmap}
$$
    K_j = \max_i \curle{ i: X_i \leq Y_{j+1} }_{i=1}^{N}
$$

Given a real number $z\in [X_0,X_N)$, the operations $j=f(z)$ and $t=K_j$, identify respectively the segment $[Y_{j},Y_{j+1})$ containing $z$ and an associated element $X_t$. Because of \eqref{eq:minh}, it is guaranteed that the segment $[Y_j,Y_{j+1})$ is smaller than the smallest segment $\min_i\curle{[X_{i},X_{i+1})}_{i=0}^{N-1}$, therefore it can contain at most one single element of the array $X$. This means that either $Y_j \leq X_t < Y_{j+1}$, implying that it could be either $z < X_t$ or $z \geq X_t$, or $X_t<Y_j$, implying that $z<X_t$. In both cases, the sought index $i$ is either $t$, if $z<X_t$, or $t-1$ otherwise.

\setlength{\unitlength}{2mm}
\begin{figure}[h]
\centering
\begin{picture}(50,16)
\linethickness{0.4mm}
\put(0,3){\line(1,0){50}}
\put(0,12){\line(1,0){50}}
 
\linethickness{0.05mm}
 
\multiput(0,2)(5,0){11} {\line(0,1){2}}
\put(-0.5,0){$Y_0$}
\put(4.5,0){$Y_1$}
\put(9.5,0){$Y_2$}
\put(14.5,0){$Y_3$}
\put(19.5,0){$Y_4$}
\put(24.5,0){$Y_5$}
\put(29.5,0){$Y_6$}
\put(34.5,0){$Y_7$}
\put(39.5,0){$Y_8$}
\put(44.5,0){$Y_9$}
\put(49.5,0){$Y_{10}$}
 
\put(0,11)  {\line(0,1){2}}
\put(6,11)  {\line(0,1){2}}
\put(15,11) {\line(0,1){2}}
\put(23,11) {\line(0,1){2}}
\put(31,11) {\line(0,1){2}}
\put(37,11) {\line(0,1){2}}
\put(50,11) {\line(0,1){2}}
\put(-0.5,14){$X_0$}
\put(5.5,14) {$X_1$}
\put(14.5,14){$X_2$}
\put(22.5,14){$X_3$}
\put(30.5,14){$X_4$}
\put(36.5,14){$X_5$}
\put(49.5,14){$X_6$}
 
\put(2.5,4) {\vector(-1,4){1.8}}
\put(7.6,4) {\vector(-1,4){1.6}}
\put(12.5,4){\vector(1,4){1.8}}
\put(17.5,4){\vector(-1,4){1.8}}
\put(23.2,4){\vector(0,2){6}}
\put(27.5,4){\vector(-1,2){3.2}}
\put(32.5,4){\vector(-1,4){1.5}}
\put(37.3,4){\vector(0,1){6}}
\put(42.5,4){\vector(-2,3){3.9}}
\put(47.7,4){\vector(1,4){1.6}}
 \end{picture}
\caption{Mapping of segments $\curle{[Y_j,Y_{j+1})}_{j=0}^R$ to elements of the array $X$}
\label{pic:directmap}
\end{figure}
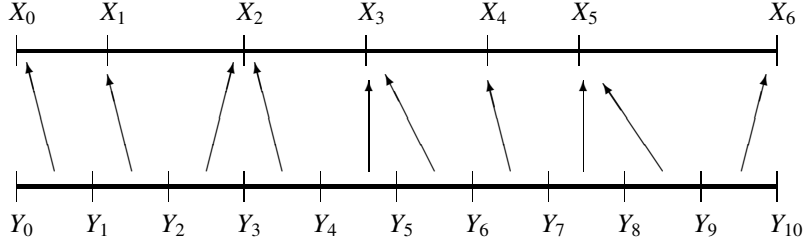

\subsection{Dealing with Floating Point Rounding Errors}
\label{sec:rounding}

The use of floating point arithmetic introduces a number of flaws in the procedure for the determination of $H$ and $R$ proposed in section \ref{sec:constructf}. 
Some examples where computations fail are illustrated below.
\begin{itemize}
	\item Given the partition $X=\{-10^{9}, 0, 1\}$ in single precision, $(X_2-X_0)$ gets rounded to $10^9$ and it is effectively indistinguishable from $(X_1-X_0)$. Given function \eqref{eq:function}, regardless of the choice of $H$, it is impossible for property \eqref{eq:f-monotonic} to hold.
	\item Given the array $X=\{0, 1.42\!\cdot\! 10^{-45} ,1\}$ in single precision, equation \eqref{eq:rtheory} overflows and yields $R = +\infty$.
\end{itemize}

Before continuing the discussion, it is useful to introduce some notation and well known facts about floating point numbers (for a more comprehensive discussion on this topic see \cite{Johnson1982} or \cite{Goldberg1991}).

\subsubsection{Facts and Notations about Floating Point Numbers}
Floating point numbers are a finite subset of rationale numbers and in a floating point numerical system any real numbers are approximated with their closest floating point number.

The result of an algebraic operation might not be a floating point number even if its operands are floating point numbers and it is therefore affected by a rounding error.

Amongst real positive numbers, only those bounded in a certain interval can be approximated with floating point numbers. Numbers too small or too big respectively underflow to zero or overflow to the abstract concept of $+\infty$.

Let: 
\begin{itemize}
\item $m(z)$ be the floating point approximation of a real number $z$
\item $\phi(x)$ be the smallest floating point number greater than the floating point number $x$
\item $\epsilon$ be the \textit{round-off error} associated with a given floating point representation, which is $\epsilon=2^{-24}$ in single precision and $\epsilon=2^{-53}$ in double precision
\end{itemize}

The relative rounding error made when a real number $x$ is approximated by the floating point number $m(x)$ is bounded by
\begin{align}
\label{eq:relerror}
1-\epsilon \leq \frac{m(x)}{x} \leq 1+\epsilon
\end{align}

\subsubsection{Feasibility Conditions}
The algebraic operations involved in the computation of \eqref{eq:function} and \eqref{eq:rhtheory} are in general affected by rounding errors, therefore it is not guaranteed that the value of $H$ they produce satisfies numerically condition \eqref{eq:fcond}.
Taking into account rounding errors, condition \eqref{eq:fcond} becomes
\begin{align}
\label{eq:cond1}
	 \floor{ m(H\, m(X_{i+1}-X_0)) } >  \floor{ m(H\, m(X_{i}-X_0)) }, \quad i=0\dots N-1
\end{align}
A necessary condition for \eqref{eq:cond1} to hold is that the results of all arithmetic subtractions involved must be numerically distinguishable, i.e. strictly increasing:
\begin{align}
\label{eq:distinguishable-exact}
	m(X_{i+1}-X_0) > m(X_{i}-X_0),   \quad i=0\dots N-1
\end{align}
Using \eqref{eq:relerror} and taking repeatedly a minorant of the left hand side and a majorant of the right hand side, condition \eqref{eq:distinguishable-exact} becomes
\begin{align}
	 (X_{i+1}-X_0)(1-\epsilon) &> (X_{i}-X_0)(1+\epsilon) ,   &&\quad i=0\dots N-1 \notag\\
	X_{i+1}-X_i &> (X_{i+1}+X_{i}-2X_0)\,\epsilon ,   &&\quad i=0\dots N-1 \notag \\
	\min_i\curle{X_{i+1}-X_i}_{i=0}^{N-1} &> (2X_N-2X_0)\,\epsilon  \notag \\
\label{eq:distinguishable-approx}
	 \dfrac{ \min_i\curle{X_{i+1}-X_i}_{i=0}^{N-1} }{X_{N}-X_0} &>2\epsilon
\end{align}

Assuming condition \eqref{eq:distinguishable-exact} holds, condition \eqref{eq:cond1} can be rewritten as a function of the rounded interval lengths $\curle{D_i=m(X_i-X_0)}_{i=0}^N$
\begin{align}
\label{eq:cond2}
	 \floor{ m(H\, D_{i+1}) } >  \floor{ m(H\, D_i) }, \quad i=0\dots N-1
\end{align}
the truncation operation can be resolved writing a more restrictive set of inequalities, 
\begin{align}
\label{eq:cond3}
	 m(H\, D_{i+1}) >  m(H\, D_i) + 1, \quad i=0\dots N-1
\end{align}
and the theoretical lower bound \eqref{eq:minh} can be approximated numerically as 
\begin{align}
\label{eq:minhnumeric}
	H > \bar{H}
		= m\left(\frac{1}{\min_i\curle{m(D_{i+1}-D_{i})}_{i=0}^{N-1}}\right)
\end{align}

The size of the array $K$ is $R=1+\floor{H\, D_N}$. In order to avoid numerical overflows, it must be $R < 2^{Q}$, where $Q$ is the number of bits used for the index returned by function \eqref{eq:function}. For function \eqref{eq:function} to be efficiently vectorizable, it is desirable that $Q=32$ or $Q\in\curle{32,64}$ depending if the array $X$ is in single precision or in double precision. Note that $Q$ is not necessarily the same as the number of bits used to represent elements of the array $K$, which is discussed in section \ref{sec:memory}. This requires that
\begin{align}
\label{eq:overflowexact}
m(H\, D_N) < 2^Q
\end{align}
Ignoring rounding errors, which might cause $H$ to be slightly larger than the theoretical lower bound \eqref{eq:minh}, this limitation can be approximately expressed in terms of the layout of the original array $X$ as
\begin{align}
\label{eq:overflowapprox}
\frac{X_N-X_0}{\min_i\curle{X_{i+1}-X_i}_{i=0}^{N-1}}< 2^Q
\end{align}
and the approximate feasibility conditions \eqref{eq:distinguishable-approx} and \eqref{eq:overflowapprox} can be combined in one single expression
\begin{equation}
\label{eq:feasibility}
\frac{\min_i\curle{X_{i+1}-X_i}_{i=0}^{N-1}}{X_N-X_0} 
    > \max \curle{2^{-Q}, 2\epsilon} 
    = \left\{ \begin{array}{l}
 	   2^{-23},\text{   in single precision, with }Q=32 \\
   	   2^{-32},\text{   in double precision, with }Q=32 \\
   	   2^{-51},\text{   in double precision, with }Q=64
    \end{array}\right.
\end{equation}
which approximately defines the family of arrays $X$ where the method is applicable.

Note that condition \eqref{eq:feasibility} is purely theoretical and cannot be verified exactly because of rounding errors.
It is however easy and inexpensive, given a value of $H$, to verify directly the original conditions \eqref{eq:cond2} and \eqref{eq:overflowexact}.

\subsubsection{Computation of a Feasible $H$}
\label{sec:feasibleh}
Both the value $\bar{H}$ computed in \eqref{eq:minhnumeric} and the evaluation of function \eqref{eq:function} are affected by rounding errors, therefore there is no guarantee that by choosing $H=\nextfloat{\bar{H}}$ condition $\eqref{eq:cond2}$ is satisfied. Should it not be, a larger value of $H$ is needed.

As discussed in section \ref{sec:constructf}, it is desirable for $H$ to be as small as possible. That poses the difficult question of \textit{how much larger should $H$ be?} There is no obvious answer and in the sequel a feasible value of $H$ is computed numerically, by increasing $H$ in small amounts in a trial and error iterative process. In brief, after a value for $H$ is chosen, the following two steps happen in a loop: condition $\eqref{eq:cond2}$ is tested for all $i$ and, if it does not hold, $H$ is progressively increased according with some growth strategy. 

A possible growth strategy for $H$ is described in algorithm \ref{alg:pseudo-computeH}, which increments $H$ adding terms of exponentially increasing size, until a feasible value is found. 
\begin{algorithm}[h]
	\caption{Computation of $H$ and $R$ (pseudo-code)}
	\label{alg:pseudo-computeH}
	\begin{algorithmic}
		\Function {ComputeHR}{\fin $\curle{X_i}_{i=0}^{N}$, $Q$ \fout $H$, $R$}
			\State	$H \leftarrow \nextfloat{\bar{H}}$ \Comment{Initialize $H$ strictly larger than the approximate lower bound \eqref{eq:minhnumeric}}
			\State {$D_{N} \leftarrow X_{N}-X_0$}
			\If {$\floor{H\,D_N \geq 2^Q}$} \Comment {Check for overflow verifying condition \eqref{eq:overflowexact}}
				\State {\textbf{ERROR}: overflow, problem unfeasible}
			\EndIf
	     	\State {$P\leftarrow\nextfloat{H}-H$} \Comment{Define a growth term $P$}
			\For {$i=1 \dots N$}
				\State {$D_{i-1} \leftarrow X_{i-1}-X_0$}
				\State {$D_{i}\;\;\; \leftarrow X_{i}\;\;\;-X_0$}
					\If {$D_{i-1}=D_{i}$} \Comment {Verify that the sequence $D_i$ is strictly increasing}
				\State {\textbf{ERROR}: $D_i$ are not strictly increasing, problem unfeasible}
				\EndIf
				\While {$\floor{H \, D_{i-1}}=\floor{H \, D_{i}}$} \Comment {Check if condition \eqref{eq:cond2} is satisfied}
					\State {$H \leftarrow H+P$} \Comment{Increase $H$}
					\If {$\floor{H\,D_N \geq 2^Q}$} \Comment Check for overflow verifying condition \eqref{eq:overflowexact}
						\State {\textbf{ERROR}: overflow, problem unfeasible}
					\EndIf 
					\State {$P \leftarrow 2P$} \Comment {Double the growth term $P$}
				\EndWhile
			\EndFor
			\State $R \leftarrow \floor{H\, D_N}$ \Comment {Compute $R$}
		\EndFunction
	\end{algorithmic}
\end{algorithm}

A minor modification of algorithm \ref{alg:pseudo-computeH} would easily allow to also track the last unfeasible value for $H$, thus defining an interval $[H_{unfeasible},H_{feasible}]$ which is known to be unfeasible at its left extreme and feasible at its right extreme. As a result instead of simply taking $H=H_{feasible}$, the solution could be refined further by bisecting this interval. This has not been implemented in this paper, as in all scenarios tested in section \ref{sec:setuptests}, $H$ is extremely rarely increased, and, when it is, the increment is negligible in relative terms.

\subsection{Memory Cost}
\label{sec:memory}
Elements of the array $K$ must have a size $B$ in bytes sufficiently large to store the largest index of the array $X$, i.e. $N$. To allow efficient memory manipulation, it should be $B\in \curle{1,2,4,8}$, i.e.
$$
B = \min\curle{\,b \in \curle{1,2,4,8}: 2^{8b} \geq N\,}
$$
Algorithm \ref{alg:direct} requires the allocation of $(R+1)\, B$ bytes. The size $R$ of the array $K$, apart from numerical related considerations, is approximately \eqref{eq:rtheory}, therefore, the total memory cost in bytes can be estimated with a quite high degree of accuracy as
$$
	MemoryCost \approx \ceil{\frac{X_N-X_0}{\min_i\curle{X_{i+1}-X_i}_{i=1}^N}} B
$$

\subsection{Initial Setup Cost}
\label{sec:setupcost}
The initial setup is divided in two parts, the computation of $H$ performed in algorithm \eqref{alg:pseudo-computeH} and the initialization of the array $K$ carried out in algorithm \eqref{alg:initk}.

The latter has computational complexity $O(N + \alpha R)$, as it requires evaluation of function \eqref{eq:function} for all elements of the array $X$ and an assignment for all elements of array $K$. The constant of proportionality $\alpha$ reflects the fact that the two operations do not have the same cost and it is $\alpha \ll 1$.

It is more difficult to estimate precisely the computational complexity associated with the computation of $H$, as it is depends on the number of iterations of the inner loop in algorithm \eqref{alg:pseudo-computeH}, which increases $H$ if needed. In theory it is of order $O(N\,(1+T))$, where $T$ is the average number of iterations in the inner loop. In practice, experimental results in section \eqref{sec:setuptests} show that $T\approx 0$, so the overall complexity is approximately $O(N)$.

The last component of the setup cost is the memory allocation of the array $K$, which is dependent on the programming language, the operating system and the memory allocation strategy used.

In section \eqref{sec:setuptests} some test results for the total setup cost, inclusive of memory allocation using the default \textit{gcc} heap allocator, are reported.
 
\subsection{Relaxing Limitations and Reducing Memory Usage}
\label{sec:bucketpairs}
It is possible to reduce the memory consumed by the index and mitigate limitations by relaxing the requirements of properties \eqref{eq:f-monotonic} as follows
\begin{align}
\label{eq:f-monotonic-2}
f(X_{i+2}) > f(X_i), &&\quad  i=0 \dots N-2
\end{align}
The definition of the index $K$ needs to be generalized as:
\begin{align}
\label{eq:kgeneral}
	K_j = \max\{i: f(X_i) \leq j\}
\end{align}
From definition \eqref{eq:kgeneral}, it follows the properties
\begin{subequations}
\label{eq:kgeneralprop}
\begin{align}
\label{eq:kgeneralprop1}
    a=f(X_{i})<f(X_{i+1})=b &\implies K_a=i \\
\label{eq:kgeneralprop2}    
    a=f(X_{i})=f(X_{i+1})=b &\implies K_a=K_b=i+1
\end{align}
\end{subequations}
Note that the modified definition of the array $K$ does not require any change to algorithm \eqref{alg:pseudo-computeH}. \\

Given a real number $z\in[X_{i},X_{i+1}\,)$, the sought index $i$ can be resolved as
\begin{equation}
\label{eq:index2}
	i=K_j-I(z<X_t)-I(z<X_{t-1})
\end{equation}
where $j=f(z)$ and $I(w)$ is the indicator function
$$
	I(w)=\left\{\begin{array}{ll}
		1, &\text{if $w$ is $true$} \\
		0, &\text{otherwise}
	\end{array}\right.
$$
The proof of statement \eqref{eq:index2} is more complicated than in the case discussed in section \ref{sec:directmethod}.
Property \eqref{eq:monotonic} of function $f$ guarantees that
$$
 	a=f(X_{i}) \leq j=f(z) \leq f(X_{i+1})=b,
$$
which implies
\begin{equation}
\label{eq:indexcond}
K_a \leq t=K_j \leq K_b,
\end{equation}
There are three possible scenarios:
\begin{enumerate}
	\item if $f(X_{i})<f(X_{i+1})<f(X_{i+2})$, from property \eqref{eq:kgeneralprop1} it follows that $K_a=i$, $K_b=i+1$, and \eqref{eq:indexcond} implies $t \in \curle{i,i+1}$
	\item if $f(X_{i})<f(X_{i+1}) \leq f(X_{i+2})$, from properties \eqref{eq:kgeneralprop} it follows that $K_a=i$, $K_b=i+2$, and \eqref{eq:indexcond} implies $t \in \curle{i,i+1,i+2}$
	\item if $f(X_{i})\leq f(X_{i+1}) < f(X_{i+2})$, from properties \eqref{eq:kgeneralprop} it follows that $K_a=i+1$, $K_b=i+2$, and \eqref{eq:indexcond} implies $t \in \curle{i+1,i+2}$
\end{enumerate}
In all cases, $t$ can only have one of the values in the set $\curle{i,i+1,i+2}$:
\begin{itemize}
	\item if $t=i$, equation \eqref{eq:index2} yields: $i-I(z<X_i)-I(z<X_{i-1})=i-0-0=i$
	\item if $t=i+1$, equation \eqref{eq:index2} yields: $i+1-I(z<X_{i+1})-I(z<X_{i})=i+1-1-0=i$
	\item if $t=i+2$, equation \eqref{eq:index2} yields: $i+2-I(z<X_{i+2})-I(z<X_{i+1})=i+1-1-1=i$
\end{itemize}
which completes the proof. \\

Ignoring rounding errors, condition \eqref{eq:fcond9} becomes
\begin{align}
\label{eq:fcond10}
	H\,(X_{i+2}-X_0) > 1+ H\,(X_i-X_0), \quad  i=0 \dots N-1
\end{align}
yielding a smaller lower bound on $H$
\begin{align}
\label{eq:fcond11}
	H > \frac{1}{\min_i\curle{X_{i+2}-X_i}_{i=0}^{N-2}}
\end{align}
which results in a smaller value of $R$ and therefore requires a smaller storage space for the auxiliary array $K$. Note that algorithm \ref{alg:pseudo-computeH} needs to be modified accordingly. \\

Limitations are also mitigated as condition \eqref{eq:feasibility} becomes
\begin{equation}
\frac{\min_i\curle{X_{i+2}-X_i}_{i=0}^{N-2}}{X_N-X_0} 
> \max \curle{2^{-Q}, 2\epsilon} 
= \left\{ \begin{array}{l}
2^{-23},\text{   in single precision, with }Q=32 \\
2^{-32},\text{   in double precision, with }Q=32 \\
2^{-51},\text{   in double precision, with }Q=64
\end{array}\right.
\end{equation}
which are clearly less restrictive. \\

Summarizing, given the function $f$ and the array $K$, the search procedure is described in algorithm \ref{alg:direct-2}. The algorithm is very similar to the one previously described, with the difference that the number $z$ needs to be compared against $X_t$ and $X_{t-1}$, with consequent performance degradation, and the index $i$ could be accordingly decreased by 1 or 2. The second conditional assignment could be made contingent on the first condition $z<X_t$, but in this case the first condition would become a genuine code flow branch, rather than a simple conditional assignment, hence it is preferable to check the two conditions independently. The downside is that, if $t=0$, this causes an access to the array element $X_{-1}$, therefore the array $X$ need to be padded to the left with one extra element containing the value $X_0$. This does not affect the correctness of the algorithm because the condition $z<X_0$ is always false.

\begin{algorithm}[ht]
	\caption{Direct Search Gap2 (scalar problem)}
	\label{alg:direct-2}
	\begin{algorithmic}
		\Function {\DirectGapName}{\fin  $z$, $\curle{X_i}_{i=0}^{N}$, $\curle{K_j}_{j=0}^{R}$, $H$, \fout $i$}
		\State $j \leftarrow \floor{H\, (z-X_0)}$
		\State $i \leftarrow K_j$
		\If {$z<X_{t}$}
		\State $i \leftarrow i-1$ \Comment {conditional assignment}
		\EndIf
		\If {$z<X_{t-1}$}
		\State $i \leftarrow i-1$ \Comment {conditional assignment}
		\EndIf
		\EndFunction
	\end{algorithmic}
\end{algorithm}

Note that this same idea could be pushed even further, generalizing condition \eqref{eq:f-monotonic} to
\begin{align}
\label{eq:f-monotonic-n}
f(X_{i+q}) > f(X_i), &&\quad  i=0 \dots N-q
\end{align}
which requires comparisons of $z$ against $X_t, X_{t-1}, \dots, X_{t-q+1}$.

If given a memory budget, this would allow to construct an optimal solution deriving the minimal number of $q$ which achieves it.

\subsection{Cache Efficient Implementation}
In algorithm \ref{alg:directcache} the following two steps always occur in close sequence:
\begin{enumerate}
	\item the index $t=K_j$ is read from the array $K$
	\item the element $X_t$ is read from the array $X$
\end{enumerate}
This suggests a reorganization of the data in memory so that $K_j$ and $X_t$ are stored contiguously in memory and part of the same cache line \footnote{a typical cache line is 64 bytes long and starts at memory addresses which are a multiple of 64}. In this way, whenever $K_j$ is fetched, $X_t$ is loaded in cache at the same time and it is already available when it is needed for the execution of the next instruction, thus reducing the number of \textit{cache misses}.

This is easily achieved by modifying the data type stored in the array $K$ to be the pair of values $(K_j,X_t)$, appropriately padded so that its total memory storage requirement is either 8 or 16 bytes, to guarantee good cache line alignment. For instance, working in single precision with 32-bit indices, the pair requires exactly 8 bytes, while working in double precision with 32-bit indices, the pair requires 12 bytes and an extra 4 bytes padding needs to be appropriately inserted.

Another advantage of this layout is that, since $K_j$ and $X_t$ are stored in contiguous aligned memory, they can be retrieved with a single aligned load instruction.

Obviously this algorithm has higher memory requirements than algorithm \ref{alg:direct}, as elements of the array $X$ are stored multiple times, plus some of the storage space might be wasted for padding.

The pseudo-code is illustrated in algorithm \ref{alg:directcache}.

\begin{algorithm}
	\caption{Direct Search Cache Friendly (scalar problem)}
	\label{alg:directcache}
	\begin{algorithmic}
		\Function {\DirectCacheName}{\fin $z$, $\curle{X_i}_{i=0}^{N}$, $\curle{K_j}_{j=0}^{R}$, $H$ \fout $i$}
		\State $j \leftarrow \floor{ H \, (z-X_0) }$
		\State {$(i,x) \leftarrow K_j$} \Comment {$K_j$ contains the pair of values $(i,X_i)$}
		\If {$z<x$}
			\State $i \leftarrow i-1$ \Comment {conditional assignment}
		\EndIf
		\EndFunction
	\end{algorithmic}
\end{algorithm}

\subsection{Fused Multiply Add}
The function \eqref{eq:function} proposed in \ref{sec:constructf} can be trivially re-factored to take advantage of \textit{fused multiply add} instructions available on modern CPUs. Let $W=H\, X_0$, function \eqref{eq:function} can be rewritten as a multiplication followed by a subtraction
\begin{equation}
\label{eq:fma}
   f(z) =\floor{H \, z-W}
\end{equation}
The reordering of algebraic operations modifies the rounding error and it is important for all phases of the algorithm, i.e. the computation of $H$, the initialization of the array $K$ and the search queries, to consistently use the function in form \eqref{eq:fma}. Also the empirical feasibility conditions \eqref{eq:cond2} and \eqref{eq:overflowexact} need to be re-factored accordingly:
\begin{align*}
	\floor{m(H\,m(X_{i+1}-W))} &> \floor{m(H\,m(X_{i}-W))},   \quad i=0\dots N-1 \\
	m(H\,m(X_{N}-W)) &< 2^Q
\end{align*}

\section{Vectorial Implementation}
\label{sec:vectorialimpl}

A vectorial implementation for the algorithms discussed is obtained by replacing each instruction with its SIMD equivalent and iterating on all elements of the array $Z$ in steps of $d$ elements, where $d$ is the SIMD bandwidth, i.e. the number of floating point numbers which can be processed simultaneously.

Conditional assignment operations are available in the SIMD instruction set, but in some cases it is possible to do even better by exploiting the fact that comparison operations set the result register to either an \textit{all-zeros} or an \textit{all-one} bit mask, permitting resolution of the conditional assignments with integer arithmetic operations, which are faster. This is discussed individually for each algorithm when the opportunity arises.

While when writing SIMD code using \textit{intrinsics}\footnote{url: \url{https://software.intel.com/sites/landingpage/IntrinsicsGuide/}} the programmer has a good degree of control on how the conditional assignments are translated in assembler, when writing plain C code it is the compiler which decides. Compilers in general do not like conditional assignment instructions because they are more expensive than regular assignment instructions, and, if the branch is predictable, the cost of the branch is negligible. This is not the case for \textit{binary search}, where the branch is unpredictable and the use of conditional assignment instructions is always the optimal choice, but the compiler cannot know that. Different compilers or even different versions of the same compiler may decide whether to use or not conditional instructions in correspondence of the same C code, so care must be taken to inspect the assembler code generated and, if needed, recourse to the use of assembler in-line.

One of the ingredients necessary for vectorization are \textit{gather} operations, which consist of reading from multiple non contiguous memory locations of the array $X$.
In the SSE instruction set there are no \textit{gather} instructions, so they must be emulated extracting data from the SIMD register, performing the memory accesses with scalar operations and inserting back the results in a SIMD register, hence perfect vectorization is not achievable.
In the AVX2 instruction set there are \textit{gather} instructions available, but, as shown in section \ref{sec:testbinvec}, they bring only a modest improvement, because on the hardware used for testing (Intel \textit{Haswell} architecture) they are implemented by emulation, as explained in the Intel Optimization Manual (\cite{Intel2016}, section 11.16.4). The problem is exacerbated even more for algorithms of the \textit{direct} family, which require two \textit{gather} operations in sequence and their use is always detrimental.

\section{Test Results}
\label{sec:results}

\subsection{Test Harness}
This section provides with some general information about the test harness.

\subsubsection{Source Code, Compiler and Hardware}
The test harness used to produce performance figures is written in C++. It uses SIMD instructions and occasionally assembler in-line. The source code is freely available from \textit{github}\footnote{url: \url{https://github.com/fabiocannizzo/fastbinarysearch.git}}.
It has been compiled with \textit{gcc 6.3.0}\footnote{command line: g++ -std=c++11 -mavx2 -mfma -O3} for \textit{Cygwin}\footnote{Cygwin is "a collection of GNU and Open Source tools which provide functionality similar to a Linux distribution on Windows" (url: \url{https://www.cygwin.com/})} 64 bits on a machine running \textit{Windows 7}.
The target machine has 32Gb RAM, 2 CPUs \textit{Intel(R) Xeon(TM)} \textit{E5-2620 v3} @2.40GHz (\textit{Haswell} architecture), each with 6 cores supporting AVX-2 and FMA instruction sets, 32Kb L1 cache, 256Kb L2 cache, 15Mb L3 cache.

\subsubsection{Data Types}
All tests results presented in this section are performed for arrays $X$ in both \textit{single} and \textit{double} precision. The index $I$ is always represented with 32-bits \textit{unsigned integers}.
Specific to algorithms of the \textit{Direct} family discussed in section \ref{sec:directmethod}, for simplicity of implementation, the indices generated by function \eqref{eq:function} and the elements of the array \eqref{eq:indexdef}  are always 32-bits \textit{unsigned integers}, regardless of the size of the arrays $K$ and $X$, which might allow the use of a smaller data type.

\subsubsection{Array $Z$ Layout}
\label{sec:arrayz}
Arrays $Z$ are always memory aligned on 64-byte boundaries. Unless otherwise specified their elements are randomly extracted with equal probability from the set of mid points $\curle{(X_i+X_{i+1})/2}_{i=0}^{N-1}$. This procedure assures that every sub-interval of the array $X$ is selected with equal probability. Size is fixed at $M=2048$, which is a value small enough to allow the entire test workspace to fit in L1 cache when the array $X$ is small; large enough to sample sufficiently well the array $X$, thus causing many cache misses, when $X$ is very large; and large enough to effectively disable the predictive capabilities of the branch prediction algorithm, as discussed in section \ref{sec:testpredict}.

\subsubsection{Array $X$ Layout}
\label{sec:arrayx} 
Arrays $X$ are composed of sub intervals of random length $\Delta_i=(X_{i+1}-X_i)$ sampled from a uniform distribution $U[1,5]$. This particular layout is chosen merely to assure that the feasibility conditions \eqref{eq:feasibility} for algorithm \ref{alg:direct} are satisfied even for the largest size of the array $X$ tested ($N \approx 2^{20}$). In fact the most restrictive of conditions \eqref{eq:feasibility} is
\begin{align*}
\frac{\min_i\curle{ X_{i+1}-X_{i} }_{i=0}^{N-1}}{X_N-X_0}
= \frac{\min_i\curle{ \Delta_i }_{i=0}^{N-1}}{\sum_{i=0}^{N-1}\Delta_i}
\geq \frac{1}{4\cdot 2^{20}}=2\cdot 2^{-23} > 2^{-23}
\end{align*}


Alternative layouts could be tested, which might lead to partitions either feasible or unfeasible with respect to algorithms of the \textit{direct} family (algorithms \ref{alg:direct}, \ref{alg:direct-2} and \ref{alg:directcache}), but this would make no general difference for the performance measurements carried out, except perhaps for considerations associated with cache memory or with effectiveness of branch prediction.

\subsubsection{Performance Measurement Tests}
In sections \ref{sec:testpredict} to \ref{sec:testFMA} the throughput of various algorithms is compared. Results are reported in millions of searches per second.

It is challenging to get accurate and reproducible time measurements on a micro processor for various reasons, like the co-existence of multiple processes, their interference in physical memory usage and the accuracy of the functions used to measure the time. Here measurements are carried out with the procedure described in algorithm \ref{alg:harness}. The function to be measured is run $R$ times without stopping or resetting the chronometer, then the average time per search is returned, thus mitigating measurement noise. This is all repeated and averaged for $G$ different data sets. The number of data sets is always fixed at $G=100$ and the number of repetitions starts from $R=20000$ and is rescaled down as the size of the array $X$ grows and computations get slower, thus requiring less repetitions to obtain results of equivalent accuracy.

\begin{algorithm}[ht]
	\caption{Test Harness (throughput test)}
	\label{alg:harness}
	\begin{algorithmic}
		\Function {TestHarness}{\fin $N$, $M$, $R$, $G$, $\curle{A_i}_{a=1}^k$ \fout $\curle{T_a}_{a=1}^k$}
  			\For {$g=1 \dots G$}  \Comment{repeat over many different arrays $X$ and $Z$}
				\State {Generate random array $X$ of size $N$}
        		\State {Generate random array $Z$ of size $M$}
                \For {$a=1 \dots k$} \Comment{test all algorithms with the same arrays $X$ and $Z$}
	                \State {Setup algorithm $A_a$}
					\State {$tStart \leftarrow GetSystemTime$}
					\For {$r=1 \dots R$}  \Comment{repeat $R$ times to mitigate time measurement noise}
						\State {$Indices \leftarrow A_a(Z)$} \Comment{resolve indices (scalar or vectorial problem)}
					\EndFor
					\State {$tFinal \leftarrow GetSystemTime$}
					\State {$W_{a,g} = R \cdot M\,/\,(tFinal-tStart)$} \Comment{throughput: algorithm $A_a$, data set $g$}
				\EndFor
			\EndFor
			\State {$T_a=(1/G)\sum_{g=1}^G \!W_{a,g} $}  \Comment{average throughput: algorithm $A_a$}
		\EndFunction
	\end{algorithmic}
\end{algorithm}

\subsubsection{Scalar vs Vectorial Problem}
The scalar problem is solved iterating over elements of the array $Z$ one by one. In-lining of the search function is not allowed, thus preventing the compiler from optimizing across multiple function calls.

The vectorial problem is solved by iterating over elements of the array $Z$ in steps of $d$ elements, where $d$ is the SIMD bandwidth. In-lining of the search function is allowed, although it only leads to a marginal improvement.

\subsubsection{Threading Model}
All tests are run on a single core in single-threaded mode. Multiple cores parallelism could be exploited by simply having multiple threads handling different search queries, or splitting the array $Z$ in blocks and having multiple threads processing them in parallel. Since every thread works in its own workspace, there is no data contention and the total throughput should scale up linearly with the number of threads, except for the additional fixed overhead associated with the management of a thread pool and potential saturation of memory I/O bandwidth. This however has not been tested.

\subsubsection{Instruction Pipelining and Explicit Memory Pre-fetching}
Khuong et al. \cite{Morin2015} reported the use of pre-fetching as beneficial. At the beginning of each iteration they pre-fetch the two possible locations needed in the next iteration. Kim et al. \cite{Kim2011} previously advised against this particular pre-fetching strategy, as it doubles memory I/O and adds extra arithmetic operations for the calculation of memory addresses. They recommend instead to pre-fetch only what is strictly needed and to pipeline multiple function calls in the vectorial problem in order to achieve a longer distance from the issue of the pre-fetch request and the time the memory is utilized.
Both these pre-fetching techniques were tested and were found to be either neutral or detrimental to performance with the CPU used for testing, hence in the tests results presented below explicit pre-fetching is not used anywhere. It is worth noting that in some cases instruction pipelining alone was found to yield up to 10\% improvement. However, this is also not used in the tests below.

\subsubsection{Algorithms Tested}

In the next sections performance tests are performed for the algorithms discussed in this paper and other algorithms proposed in the literature or current software libraries. To make results more readable friendly nicknames are assigned to each algorithm and their main traits are summarized here below.
\subsubsubsection{Algorithms with Complexity $O(log N)$}
	\begin{myitemize}
		\item \textit{\ClassicName} is is algorithm \ref{alg:naivealg}, proposed by Press et al. \cite{NRC++} and used as a baseline algorithm in this paper. It is not vectorizable, the number of iterations is not fixed and it requires a control flow branch.
		\item \textit{\ClassicModName} is a variation of algorithm \ref{alg:naivealg} briefly discussed at the end of section \ref{sec:binary}. It is branch free and vectorizable, the number of iterations is fixed and the control flow branch is replaced by two conditional assignments.
		\item \textit{\MorinBranchyName} is an implementation used as a baseline in Khuong et al. \cite{Morin2015} with a 3-way  control flow branch allowing for early exit if an exact match is found. It is not vectorizable.
		\item \textit{\LowerBoundName} is part of the STL. It contains branches and is not vectorizable. The implementation used is the one provided by gcc. 
		\item \textit{\BitSetName} is algorithm \ref{alg:binaryopt}, a variation of Pulver \cite{Pulver2011}, which does not impose limitations on $N$, but requires extra memory for padding the array $X$. It is vectorizable and requires one conditional assignment. 
		\item \textit{\BitSetNoPadName} is algorithm \ref{alg:binaryoptnopad}, a variation of algorithm \ref{alg:binaryopt}, which does not require extra memory for padding the array $X$, at the cost of introducing extra checks on the index range.		
		\item \textit{\MorinOffsetName} is a branch free variation of Knuth \cite{Knuth1997} proposed in Khuong et al. \cite{Morin2015} and discussed in section \ref{sec:offset}. It is vectorizable and requires one conditional assignment, but the use of pointers would make a vectorial implementation a bit convoluted.
		\item \textit{\ClassicOffsetName} is algorithm \ref{alg:naiveoffset}, a branch free hybrid of Knuth \cite{Knuth1997} and Khuong et al. \cite{Morin2015}. It is vectorizable and requires one conditional assignment.
		\item \textit{\EytzingerName} is algorithm \ref{alg:eytzinger}, a variation of Sanders et al. \cite{Sanders2004} discussed in section \ref{sec:eytzinger}, which does not impose limitations on $N$, but requires extra memory for padding the array $X$. It is branch free and vectorizable and requires one conditional assignment. The algorithm is designed to use efficiently cache memory, which is important with arrays $X$ of large size.
		\item \textit{\TernaryName}, \textit{\PentaryName} and \textit{\NonaryName} are specializations of Schlegel et al. \cite{kary2009} \textit{k-ary search} algorithm for some chosen values of $k$. The array $X$ is subdivided in $k$ sub-intervals at every step, yielding time complexity $O(log_k N)$, but increasing the number of comparisons performed. To compensate, the comparisons are performed in parallel using SIMD instructions, which are thus used to solve the scalar problem. The array $X$ is padded at the end with $X_N$ to reach perfect size $k^h-1$ for some positive integer $h$, as proposed in Zeuch et al. \cite{kary2014}, then it is rearranged as in a breadth first linearized tree where every node contains exactly $k-1$ interval separators. This layout is conceptually equivalent to \textit{\EytzingerName}, which can be considered as a special case of this algorithm with $k=2$, therefore it is expected to be also cache friendly. The implementations tested are chosen to match the available SIMD bandwidths, i.e. $k=3$ using SSE instructions in double precision, $k=5$ using SSE instructions in single precision or AVX instructions in double precision, and $k=9$ using AVX instructions in single precision. All memory accesses are aligned and contiguous, making memory I/O quite efficient.
	\item \textit{\MKLName} is a proprietary algorithm available in the Intel Math Kernel Library (version 2017-3). The functions used are \textit{dfsSearchCells1D} in single precision and \textit{dfdSearchCells1D} in double precision. The partition $X$ is initialized with the \textit{QUASI\_UNIFORM} hint, which yields experimental results superior to the alternatives for the particular layouts of the array $X$ used in this test. The functions are designed to solve the vectorial problem, hence numerical results reported for the scalar problem are obtained calling the vectorial function with arrays $Z$ of size $M=1$, which probably constitutes a penalizing scenario for the functions.
\end{myitemize}

The algorithms \textit{\MorinBranchyName}, \textit{\LowerBoundName} and \textit{\MorinOffsetName} by design solve a problem slightly different from the one described in section \ref{sec:definition}. They search for the smallest index $i$ such that $z \geq X_i$, i.e. for the interval such that $z \in (X_{i-1},X_i]$. To reconcile this it suffices to feed them with an array $X$ sorted in descending order and to replace everywhere the condition $z \geq X_i$ with $z \leq X_i$. Note that these code changes are cost-neutral, i.e. they do not alter the performance of the algorithm thus allowing a fair comparison. Furthermore, Khuong's original algorithm contains an extra iteration at the end to handle the case when $X \geq X_N$, which is unnecessary given the assumption that $z \in [X_0,X_N)$ and, for sake of a fair comparison, is removed.

\subsubsubsection{Algorithms with Complexity $O(1)$}
	\begin{myitemize}
		\item \textit{\DirectName} is algorithm \ref{alg:direct}.
		\item \textit{\DirectGapName} is algorithm \ref{alg:direct-2}, which uses less memory at the cost of doubling the number of comparisons.
		\item \textit{\DirectCacheName} is algorithm \ref{alg:directcache}, which optimizes memory accesses at the cost of using more memory.
		\item \textit{\DirectFMAName}, \textit{\DirectGapFMAName} and \textit{\DirectCacheFMAName} are implementations of the above algorithms using FMA instructions.
	\end{myitemize}


\subsection{Impact of Branch Elimination}
\label{sec:testpredict}
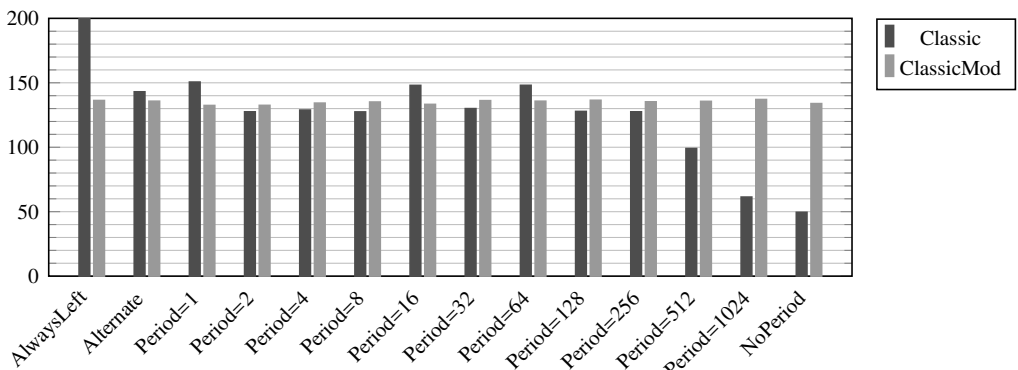
\begin{figure}[ht]
\centering
\begin{tikzpicture}
    \begin{axis}[
        width  = 0.9\textwidth,
        height = 5cm,
        major x tick style = transparent,
        ybar=2.5*\pgflinewidth,
        bar width=4pt,
        ymajorgrids = true,
        yminorgrids=true,
        minor y tick num=4,
        ylabel style = {font=\footnotesize},
		ytick={0, 50,...,300},
        symbolic x coords={AlwaysLeft,Alternate,Period=1,Period=2,Period=4,Period=8,Period=16,Period=32,Period=64,Period=128,Period=256,Period=512,Period=1024,NoPeriod},
        xtick = data,
		xticklabel style={rotate=45, anchor=east, font=\footnotesize},
        yticklabel style = {font=\footnotesize},
        scaled y ticks = false,
        enlarge x limits=0.06,
        ymin=0,
        ymax=200,
		legend pos=outer north east,
		legend style={font=\footnotesize}
    ]

  \addplot[style={black!70,fill,mark=none}]
coordinates {(Period=1,150.583) (Period=2,127.495) (Period=4,128.855) (Period=8,127.427) (Period=16,148.009) (Period=32,129.995) (Period=64,148.065) (Period=128,127.795) (Period=256,127.505) (Period=512,99.032) (Period=1024,61.3618) (NoPeriod,49.5286)  (Alternate,143.028) (AlwaysLeft,206.752) };
\addplot[style={black!40,fill,mark=none}]
coordinates {(Period=1,132.424) (Period=2,132.51) (Period=4,134.244) (Period=8,135.097) (Period=16,133.263) (Period=32,136.168) (Period=64,135.727) (Period=128,136.444) (Period=256,135.264) (Period=512,135.601) (Period=1024,137.055) (NoPeriod,133.879) (Alternate,135.728) (AlwaysLeft,136.271) };

        \legend{\ClassicName,\ClassicModName}
    \end{axis}   
\end{tikzpicture}
\vspace*{-10mm}
\caption{Throughput in millions of searches per second with array X of size $N=15$ and array $Z$ of size $M=2048$ with different layouts}
\label{fig:predictor}
\end{figure}

This test illustrates the impact of branch elimination and the effect of branch mis-predictions.
The size of the arrays $X$ and $Z$ is $N=15$ and $M=2048$, hence the workspace fits entirely in L1 cache and there are no cache miss penalties. The test compares the performance of \textit{\ClassicName} against \textit{\ClassicModName}, which is its closest branch free equivalent, for various layouts of the array $Z$. The test results are illustrated in figure \ref{fig:predictor} and are representative for both single and double precision.

\textit{Always Left}. The array $Z$ contains only one distinct value, i.e. $Z_j=X_0, \forall j$, which causes the branch in the inner loop of algorithm \ref{alg:naivealg} to be resolved always in the same direction. Branch prediction easily achieves a perfect score and \textit{\ClassicName} performs even better than \textit{\ClassicModName}, which is branch free but involves conditional assignments.

\textit{Alternate}. The array $Z$ contains only one distinct value, i.e. $Z_j=X_i, \forall j$, where $i$ is chosen so that the branch in the inner loop of \textit{\ClassicName} resolves in alternate directions at each iteration. Although the performance degradation is noticeable, the ability of the branch predictor to cope even in the presence of complex branching patterns is remarkable.

\textit{Period=$P$}. The array $Z$ is populated with a sequence of $P$ random values repeated periodically. As the length of the period $P$ becomes large, the performance of \textit{\ClassicName} decreases significantly, while the performance of \textit{\ClassicModName} is not affected.

\textit{No Period}. Since $2048$ is the length of array $Z$, this case is conceptually equivalent to \textit{Period=2048}, and approaches closely the minimum of performance for the branch prediction algorithm, which would not degrade much further for larger values of $P$.

\subsection{Algorithms of the \textit{Binary Search} Family}
\label{sec:testbinalg}
\begin{figure}[ht]
\begin{tikzpicture}
    \begin{axis}[
        width  = 1*\textwidth,
        height = 5.5cm,
        major x tick style = transparent,
        ybar=3*\pgflinewidth,
        bar width=3pt,
        ymajorgrids = true,
         yminorgrids=true,
        minor y tick num=4,
		ytick={0, 50,..., 1000},
        symbolic x coords={\ClassicName,\LowerBoundName,\MorinBranchyName,\MorinOffsetName,\ClassicModName,\BitSetNoPadName,\ClassicOffsetName,\BitSetName,\EytzingerName},
        xtick = data,
        ylabel style = {font=\footnotesize},
        yticklabel style = {font=\footnotesize},
		xticklabel style={rotate=45, anchor=east,font=\footnotesize},
        scaled y ticks = false,
        enlarge x limits=0.08,
        ymin=0,
        ymax=410,
	    legend pos=north west,
   		legend style={font=\footnotesize},
    ]

\addplot[style={black!80,fill}]
coordinates {(\EytzingerName,173.996) (\BitSetName,202.627) (\ClassicOffsetName,183.881) (\MorinOffsetName,145.068) (\BitSetNoPadName,160.195) (\ClassicModName,135.92) (\MorinBranchyName,59.4216) (\ClassicName,49.9877) (\LowerBoundName,59.3079) };
\addplot[style={black!40,fill}]
coordinates {(\EytzingerName,174.761) (\BitSetName,208.218) (\ClassicOffsetName,186.305) (\MorinOffsetName,146.761) (\BitSetNoPadName,164.728) (\ClassicModName,135.98) (\MorinBranchyName,60.0476) (\ClassicName,50.7165) (\LowerBoundName,59.8088) };
\addplot[style={black!80,fill=white,pattern color=black!80,postaction={pattern=north east lines}}]
coordinates {(\EytzingerName,366.095) (\BitSetName,374.587) (\ClassicOffsetName,369.717) (\BitSetNoPadName,364.79) (\ClassicModName,215.613) };
\addplot[style={black!40,fill=white,pattern color=black!40,postaction={pattern=north east lines}}]
coordinates {(\EytzingerName,253.338) (\BitSetName,261.189) (\ClassicOffsetName,250.586) (\BitSetNoPadName,244.75) (\ClassicModName,134.799) };
\addplot[style={black!80,fill=white,pattern color=black!80,postaction={pattern=north west lines}}]
coordinates {(\EytzingerName,401.222) (\BitSetName,399.666) (\ClassicOffsetName,399.58) (\BitSetNoPadName,392.954) (\ClassicModName,248.516) };
\addplot[style={black!40,fill=white,pattern color=black!40, postaction={pattern=north west lines}}]
coordinates {(\EytzingerName,284.288) (\BitSetName,219.844) (\ClassicOffsetName,220.262) (\BitSetNoPadName,215.914) (\ClassicModName,136.877) };

       \legend{\SingleName-\ScalarName,\DoubleName-\ScalarName,\SingleName-\SSEName,\DoubleName-\SSEName,\SingleName-\AVXName,\DoubleName-\AVXName}
    \end{axis}
\end{tikzpicture}
\vspace*{-6mm}
\caption{Throughput in millions of searches per second with array X of size $N=15$ and array $Z$ of size $M=2048$ randomly populated}
\label{fig:selectbinary}
\end{figure}
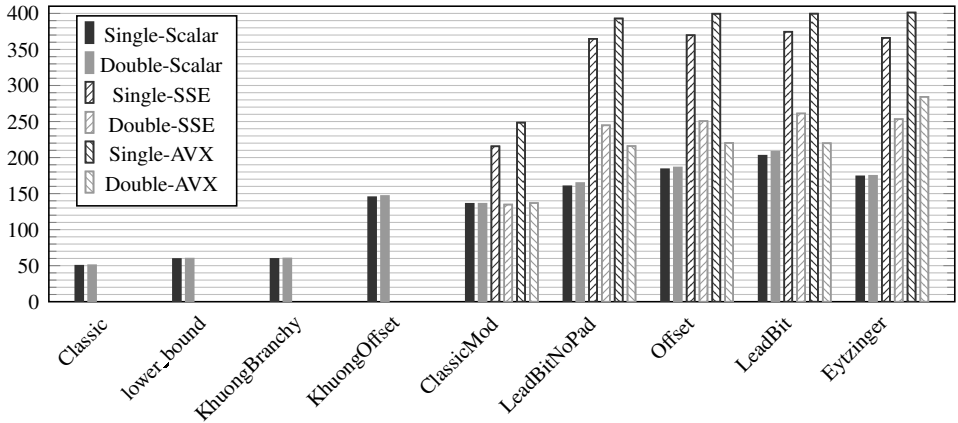

This test compares the performance of the various implementations of the \textit{binary search} algorithm with time complexity $O(log_2N)$.
The size of the arrays $X$ and $Z$ is $N=15$ and $M=2048$, hence the workspace fits entirely in L1 cache and there are no cache misses.
The array $Z$ is randomly generated as descibed in section \ref{sec:arrayz}.
A vectorial implementation is tested for all branch free algorithms, except for \textit{\MorinOffsetName}, where the implementation based on pointers would be more convoluted.
The test results are shown in table \ref{tab:results0} and illustrated in figure \ref{fig:selectbinary}.

\subsubsection{Scalar Problem Considerations}

The test results show the superior performance in the scalar problem of all branch free algorithms. Out of these the best performer is \textit{\BitSetName}, with its essential arithmetic, closely followed by \textit{\ClassicOffsetName}.

Despite the fact that the size of array $X$ is favorable to \textit{\BitSetNoPadName}, as there are zero iterations in the second more expensive loop, its more convoluted logic causes a noticeable degradation in performance with respect to \textit{\BitSetName} and \textit{\ClassicOffsetName}. Given that \textit{\ClassicOffsetName} also does not require any extra memory, it can be considered strictly superior.

\textit{\MorinOffsetName}, with its pointer arithmetic, performs just slightly worse than \textit{\ClassicOffsetName}, which has identical access pattern and can therefore be considered strictly superior.

\textit{\EytzingerName} performs slightly worse than other algorithms, which is expected due to its more complex arithmetic, but it has a different memory access pattern which promises to make a difference when the size of the array $X$ becomes very large and cache friendliness becomes important.

\textit{\ClassicModName} performs noticeably worst than other branch free algorithms, as expected due to the two conditional assignments.

Results in single and double precision are conceptually identical, as expected in the absence of memory I/O penalties on a 64 bits machine, because the cost of comparing floating point numbers in single or double precision is the same.

\subsubsection{Vectorial Problem Considerations}
\label{sec:testbinvec}
Vectorial implementations with SSE instructions show a material performance improvement over the scalar ones. In double precision the throughput is almost double, but in single precision the improvement is smaller. Perfect vectorization is not achieved because of the unavailability of \textit{gather} instructions. The throughput is proportional to $d/(\alpha + \beta\,d)$, where $\alpha$ and $\beta$ are the costs associated with vectorizable and sequential instructions and the coefficient $\beta$ changes with the data precision and the instruction set used. Therefore the improvement obtained does not simply scale up with the SIMD bandwidth. 

The AVX instruction set offers native \textit{gather} operations, but performance is generally disappointing. Throughput is nearly identical to the SSE case, slightly better in single precision and worse in double precision. This is because on Haswell CPUs the \textit{gather} instructions are implemented by emulation, i.e. memory accesses are serialized and the micro operations necessary to insert and extract from the SIMD registers are fused to different extents for different precisions, as discussed in section \ref{sec:vectorialimpl}.

\subsection{Performance Tests with Arrays $X$ of Different Size}
\label{sec:testthroughput}
Tables \ref{tab:results0}-\ref{tab:results4} show the throughput of various algorithms for arrays $X$ of size $2^h-1$, where $h=\curle{4,8,12,16,20}$, generated as described in section \ref{sec:arrayx}. The choice of these particular array sizes allows optimal memory usage for \textit{\EytzingerName} and \textit{\BitSetName}. The array $Z$ has size $M=2048$ and is randomly generated. Test results for the scalar and vectorial problems are illustrated respectively in figure \ref{fig:perf-scalar} and \ref{fig:perf-SSE}, which are extracts of tables \ref{tab:results0} to \ref{tab:results4}.

\pgfplotsset{width=6cm} 

\begin{figure}
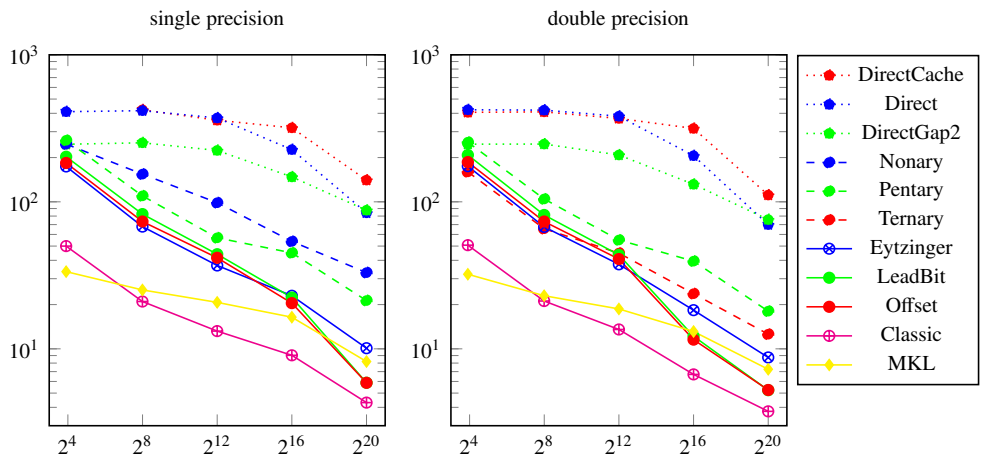
%
\begin{tabular}{@{\hskip1pt}c@{\hskip1pt} @{\hskip1pt}c@{\hskip1pt}}
		
	\begin{tikzpicture}
	\begin{loglogaxis}[
			height=6.5cm,
			title={single precision},
			xmin=8, xmax=2097152,
			ymin=3, ymax=1000,
			log basis x=2,
			log basis y=10,
			xtick={16,256,4096,65536,1048576},
			xminorticks=false,
			yminorticks=true,
			legend pos=outer north east,
			grid style=dashed,
		]
	\input{scalar-single.perfplot}
	\end{loglogaxis}
	\end{tikzpicture}
 & 
	\begin{tikzpicture}
	\begin{loglogaxis}[
			height=6.5cm,
			title={double precision},
			xmin=8, xmax=2097152,
			ymin=3, ymax=1000,
			log basis x=2,
			log basis y=10,
			xtick={16,256,4096,65536,1048576},
			xminorticks=false,
			yminorticks=true,
			legend pos=outer north east,
			grid style=dashed,
		]
	\input{scalar-double.perfplot}
	\end{loglogaxis}
	\end{tikzpicture}

\end{tabular}
\vspace*{-4mm}
\caption{Throughput for the scalar problem in millions of searches per second vs the size of the array $X$ in single and double precision (resp. left and right pane).}
\label{fig:perf-scalar}%
\end{figure}

\subsubsection{Scalar Problem Considerations}

\textit{\MKLName} performance is generally superior to \textit{\ClassicName}, but significantly inferior to any of the enhanced versions of \textit{binary search} discussed. For small arrays, it is even slower than \textit{\ClassicName}, which is not surprising considering that the function used has a vectorial signature and incurs unnecessary overheads to deal with just a scalar. Its time complexity is better than logarithmic and it is more resilient than other algorithms to cache memory related penalties for arrays $X$ of large size.

The three enhanced versions of \textit{binary search} (\textit{\BitSetName}, \textit{\ClassicOffsetName} and \textit{\EytzingerName}) for arrays of small to medium size exhibit only minor differences, with \textit{\BitSetName} being slightly faster. When the array $X$ gets large, as expected, there is a steep drop in performance, except for the cache friendly \textit{\EytzingerName}, which is resilient to that.

\textit{K-ary} search methods exhibit a performance superior to any other method with logarithmic complexity. This is not surprising as they use SIMD instructions for the solution of the scalar problem, whereas other logarithmic methods do not, except perhaps for MKL. Since their layout is comparable to the one used by \textit{\EytzingerName}, they are also not affected by the drop in performance for large arrays.

All versions of \textit{direct search} (\textit{\DirectName}, \textit{\DirectGapName}, \textit{\DirectCacheName}) run in constant time and are not affected by the size of the array $X$ until it reaches dimensions so large that cache memory effects come into play. They exhibit a throughput increase of at least one order of magnitude with respect to \textit{\ClassicName} and significantly faster than the best alternative logarithmic algorithm. This is no surprise given the difference in computational complexity. As expected, \textit{\DirectGapName} has slightly inferior performance than \textit{\DirectName}, due to the increased computation cost, however this difference disappears as the size of the array grows and cache misses becomes the bottleneck. \textit{\DirectCacheName} is superior to all others not only for large vectors, where an efficient utilization of cache memory becomes critical, but also for small vectors because it retrieves all the data it needs with a single aligned read.

\begin{figure}
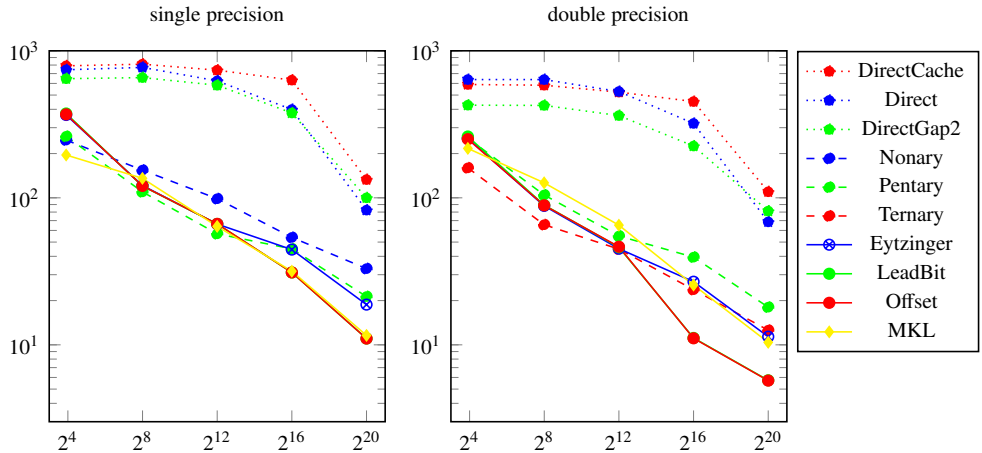
%
	\begin{tabular}{@{\hskip1pt}c@{\hskip1pt} @{\hskip1pt}c@{\hskip1pt}}
		
	\begin{tikzpicture}
	\begin{loglogaxis}[
			height=6.5cm,
			title={single precision},
			xmin=8, xmax=2097152,
			ymin=3, ymax=1000,
			log basis x=2,
			log basis y=10,
			xtick={16,256,4096,65536,1048576},
			xminorticks=false,
			yminorticks=true,
			legend pos=outer north east,
			grid style=dashed,
		]
	\input{SSE-single.perfplot}
	\end{loglogaxis}
	\end{tikzpicture}
 & 
	\begin{tikzpicture}
	\begin{loglogaxis}[
			height=6.5cm,
			title={double precision},
			xmin=8, xmax=2097152,
			ymin=3, ymax=1000,
			log basis x=2,
			log basis y=10,
			xtick={16,256,4096,65536,1048576},
			xminorticks=false,
			yminorticks=true,
			legend pos=outer north east,
			grid style=dashed,
		]
	\input{SSE-double.perfplot}
	\end{loglogaxis}
	\end{tikzpicture}

	\end{tabular}
	\vspace*{-4mm}
	\caption{Throughput for the vectorial problem with SSE instruction set in millions of searches per second vs the size of the array $X$ in single and double precision (resp. left and right pane).}
	\label{fig:perf-SSE}%
\end{figure}

\subsubsection{Vectorial Problem Considerations}

Except for the smallest size of the array $X$ ($N=15$), \textit{\MKLName} is comparable or better than the enhanced variations of \textit{binary search}. Time complexity is logarithmic and it is resilient to cache memory effects for arrays $X$ of large size.

For {K-ary search} algorithms, the results reported in figure \ref{fig:perf-SSE} are those obtained iterating on the scalar problem. This shows that the algorithms are competitive also with the vectorial implementations of the enhanced variations of \textit{binary search}.

The performances of \textit{\ClassicOffsetName} and \textit{\BitSetName} are hardly distinguishable, because the fixed cost associated with their arithmetic operations becomes negligible compared to the cost of \textit{gather} operations, as explained in section \ref{sec:testbinvec}. For small arrays \textit{\EytzingerName} is also comparable, but it becomes faster for larger arrays.

Algorithms of the \textit{direct} family produce a throughput massively superior to any other algorithm. The difference between \textit{\DirectName} and \textit{\DirectGapName} is relatively smaller than in the scalar case, as the increased number of comparisons is amortized via the use of SIMD instructions. 

\begin{figure}[ht]
	\begin{tikzpicture}
	\begin{axis}[
	width  = 0.85*\textwidth,
	height = 5.5cm,
	major x tick style = transparent,
	ybar=3*\pgflinewidth,
	bar width=3pt,
	ymajorgrids = true,
	yminorgrids=true,
	minor y tick num=3,
	ytick={0, 100,..., 1000},
	symbolic x coords={\DirectName,\DirectFMAName,\DirectCacheName,\DirectCacheFMAName,\DirectGapName,\DirectGapFMAName},
	xtick = data,
	ylabel style = {font=\footnotesize},
	xticklabel style={rotate=45, anchor=east,font=\footnotesize},
	yticklabel style = {font=\footnotesize},
	scaled y ticks = false,
	enlarge x limits=0.08,
	ymin=0,
	ymax=900,
	legend pos=outer north east,
	legend style={font=\footnotesize},
	]
	
	\addplot[style={black!80,fill}]
	coordinates {(\DirectCacheFMAName,448.793) (\DirectFMAName,440.349) (\DirectGapFMAName,267.34) (\DirectCacheName,412.61) (\DirectName,410.283) (\DirectGapName,246.33) };
	\addplot[style={black!40,fill}]
	coordinates {(\DirectCacheFMAName,414.771) (\DirectFMAName,421.873) (\DirectGapFMAName,257.316) (\DirectCacheName,406.192) (\DirectName,422.844) (\DirectGapName,246.631) };
	\addplot[style={black!80,fill=white,pattern color=black!80,postaction={pattern=north east lines}}]
	coordinates {(\DirectCacheFMAName,847.668) (\DirectFMAName,823.178) (\DirectGapFMAName,708.089) (\DirectCacheName,790.205) (\DirectName,742.861) (\DirectGapName,646.097) };
	\addplot[style={black!40,fill=white,pattern color=black!40,postaction={pattern=north east lines}}]
	coordinates {(\DirectCacheFMAName,652.559) (\DirectFMAName,695.212) (\DirectGapFMAName,453.121) (\DirectCacheName,590.195) (\DirectName,637.41) (\DirectGapName,426.751) };
	\addplot[style={black!80,fill=white,pattern color=black!80,postaction={pattern=north west lines}}]
	coordinates {(\DirectCacheFMAName,814.034) (\DirectFMAName,732.21) (\DirectGapFMAName,685.85) (\DirectCacheName,780.618) (\DirectName,705.74) (\DirectGapName,661.705) };
	\addplot[style={black!40,fill=white,pattern color=black!40,postaction={pattern=north west lines}}]
	coordinates {(\DirectCacheFMAName,668.08) (\DirectFMAName,674.138) (\DirectGapFMAName,583.28) (\DirectCacheName,625.65) (\DirectName,629.327) (\DirectGapName,549.452) };
	
	\legend {\SingleName-\ScalarName,\DoubleName-\ScalarName,\SingleName-\SSEName,\DoubleName-\SSEName,\SingleName-\AVXName,\DoubleName-\AVXName}
	\end{axis}
	
	\end{tikzpicture}
	\vspace*{-5mm}
	\caption{Throughput in millions of searches per second with array X of size $N=15$ and array $Z$ of size $M=2048$ randomly populated}
	\label{fig:selectdirect}
\end{figure}
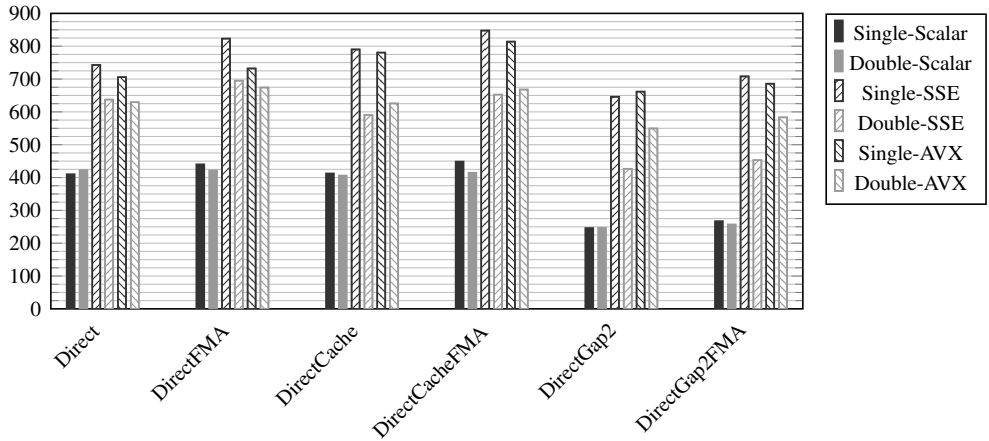

\subsection{Impact of FMA with Algorithms of the \textit{Direct} Family}
\label{sec:testFMA}
This test compares the performance of the various implementations of the \textit{direct search} algorithm.
The size of the arrays $X$ and $Z$ is $N=15$ and $M=2048$, hence the workspace fits entirely in L1 cache and there are no cache misses. 
The array $Z$ is randomly generated sampling with equal probability each segment of the array $X$.

The test results in figure \ref{fig:selectdirect}, which are extracted from table \ref{tab:results0}, show  
that for each variation of the algorithm the use of FMA instructions leads to an improvement of up to 10\%.

\subsection{Direct Search Setup Cost Test}
\label{sec:setuptests}
\subsubsection{Details}
As discussed in section \ref{sec:setupcost}, the setup cost is uncertain because of the unpredictable number of times $H$ is increased in the inner loop of \ref{alg:pseudo-computeH}. Therefore all test results in this section are described with statistical properties over populations of randomly generated arrays $X$. Measurements are repeated many times, similar to what is done in algorithm \ref{alg:harness}, to reduce measurement noise.

Table \ref{tab:results101} presents the number of times $H$ is increased in the inner loop of algorithm \ref{alg:pseudo-computeH}. A sample population of 10000 randomly generated arrays $X$ is used.

\begin{table}[ht]
	\footnotesize
	\centering
	\begin{tabular}{| c | c c c c | c c c c |}
		\cline{2-9}
		\multicolumn{1} {c|}{}  & \multicolumn{4}{c|}{\textbf{Single Precision}}  & \multicolumn{4}{c|}{\textbf{Double Precision}} \\
		\hline
		\textbf{array size} & \textbf{mean} & \textbf{min} & \textbf{max} & \textbf{stdev} & \textbf{mean} & \textbf{min} & \textbf{max} & \textbf{stdev} \\
		\hline
		\multicolumn{1}{|c|}{\textbf{15}                                    } &     0.0061 &     0.0000 &     1.0000 &     0.0779 &     0.0048 &     0.0000 &     1.0000 &     0.0691 \\
		\multicolumn{1}{|c|}{\textbf{255}                                   } &     0.0000 &     0.0000 &     0.0000 &     0.0000 &     0.0001 &     0.0000 &     1.0000 &     0.0100 \\
		\multicolumn{1}{|c|}{\textbf{4095}                                  } &     0.0000 &     0.0000 &     0.0000 &     0.0000 &     0.0000 &     0.0000 &     0.0000 &     0.0000 \\
		\multicolumn{1}{|c|}{\textbf{65535}                                 } &     0.0000 &     0.0000 &     0.0000 &     0.0000 &     0.0000 &     0.0000 &     0.0000 &     0.0000 \\
		\multicolumn{1}{|c|}{\textbf{1048575}                               } &     0.0000 &     0.0000 &     0.0000 &     0.0000 &     0.0000 &     0.0000 &     0.0000 &     0.0000 \\
		\hline
	\end{tabular}
	\caption{Number of $H$ updates}
	\label{tab:results101}
\end{table}

Table \ref{tab:results100} presents the setup cost in nanoseconds normalized by the size of the array $X$. A sample population of 1000 randomly generated arrays $X$ is used.

Multiplying the test results in table \ref{tab:results100} for the array size yields the average setup time in nanoseconds. This allows for a direct comparison with the results presented in tables  \ref{tab:results0}-\ref{tab:results4}, which are expressed in millions of searches per seconds, making it possible to express the setup costs in terms of equivalent number of searches.

\begin{table}[ht]
	\centering
	\footnotesize
	\begin{tabular}{| c | c c c c | c c c c |}
		\cline{2-9}
		\multicolumn{1} {c|}{}  & \multicolumn{4}{c|}{\textbf{Single Precision}}  & \multicolumn{4}{c|}{\textbf{Double Precision}} \\
		\hline
		\textbf{array size} & \textbf{mean} & \textbf{min} & \textbf{max} & \textbf{stdev} & \textbf{mean} & \textbf{min} & \textbf{max} & \textbf{stdev} \\
		\hline
		\multicolumn{1}{|c|}{\textbf{15}                                    } &      10.24 &       8.96 &      11.59 &       0.47 &      11.81 &      10.11 &      13.31 &       0.49 \\
		\multicolumn{1}{|c|}{\textbf{255}                                   } &       8.27 &       8.00 &       8.66 &       0.12 &       8.40 &       8.06 &       8.65 &       0.13 \\
		\multicolumn{1}{|c|}{\textbf{4095}                                  } &      13.80 &      13.54 &      13.98 &       0.09 &      13.51 &      13.30 &      13.70 &       0.07 \\
		\multicolumn{1}{|c|}{\textbf{65535}                                 } &      17.63 &      17.61 &      17.65 &       0.01 &      17.49 &      17.46 &      17.51 &       0.01 \\
		\multicolumn{1}{|c|}{\textbf{1048575}                               } &      17.67 &      17.66 &      17.69 &       0.01 &      17.61 &      17.58 &      17.66 &       0.02 \\
		\hline
	\end{tabular}
	\caption{Statistical setup cost for algorithm \ref{alg:direct} in nano seconds normalized by the array size}
	\label{tab:results100}
\end{table}

\subsubsection{Considerations}
Table \ref{tab:results101} shows that the number of times $H$ is increased in the inner loop of algorithm \ref{alg:pseudo-computeH} is independent of $N$. At most one iteration is carried out, regardless of the size of the array $X$. This implies that the initial guess \eqref{eq:minhnumeric} is acceptable in almost all cases and, when an increase is necessary, it is negligible in relative terms. Hence the approximate limiting conditions \eqref{eq:feasibility} which are derived from \eqref{eq:minhnumeric} have good general accuracy.

Table \ref{tab:results100} shows that the setup cost normalized by the size of the array $X$ initially decreases as the size of the array $X$ grows, but then it increases until it stabilizes on an asymptotic level.
The initial decrease can be explained as the effect of the fixed overhead associated with the algorithm, due for instance to the allocation of memory for the array $K$, which is material for a small array, but gets amortized over larger arrays. 
The subsequent increase can be explained by the larger amount of total memory necessary to store the arrays $X$ and $K$, which results in progressively inefficient use of the cache memory.

From these experimental results it is possible to conclude that, apart from memory I/O effects, the setup cost per element is independent of $N$, i.e. the setup has complexity $O(N)$.


\section{Conclusion}
Possible technical improvements to the classic \textit{binary search} algorithm are described. Although having the same asymptotic complexity, these are vectorizable and generally faster. Test results show that \textit{\ClassicOffsetName} is the fastest algorithm not requiring any extra storage. Otherwise \textit{\BitSetName} and \textit{\EytzingerName} are the best scalar algorithms not requiring SIMD instructions respectively for small or medium size and large size arrays. \textit{K-ary search} algorithms are a superior alternative if SIMD instructions are available.

Next a new algorithm with superior asymptotic complexity is presented. This only requires one multiplication, one subtraction and two memory accesses and can be implemented using FMA instructions. Test results using streaming SIMD extensions demonstrate that with arrays $X$ of various length randomly populated, the proposed algorithm is up to 70 times faster than the \textit{classic binary search} (see table \ref{tab:results3}). A cache friendly version of the algorithm is also proposed, which arranges the two needed memory locations in contiguous memory on the same cache line and leads to even superior performance. There are situations where the algorithm is not applicable. These are extensively discussed and can be inexpensively identified in the preliminary analysis phase, thus allowing fall-back to some of the algorithms with logarithmic complexity. Possible variations of the algorithm, which mitigate such limitations sacrificing some performance, are also proposed.

With large arrays, memory access speed is the bottleneck for all algorithms, as already found in previous studies.

When SIMD instuctions are used to solve the vectorial problem, the unavailability of effective \textit{gather} instructions on modern CPUs prevents throughput to scale up by more than a factor of 2, however this may change as technology keeps enhancing.



\pagebreak
\appendix

\section{Numerical Results}

\begin{table}[ht]
\centering
\footnotesize

\begin{tabular}{l | c c c | c c c |}
\cline{2-7}
              & \multicolumn{3}{c|}{\textbf{Single}} & \multicolumn{3}{c|}{\textbf{Double}} \\
\cline{2-7}
              & \textbf{Scalar} & \textbf{SSE-4} & \textbf{AVX-2} & \textbf{Scalar} & \textbf{SSE-4} & \textbf{AVX-2} \\
              & $\mathbf{d=1}$ & $\mathbf{d=4}$ & $\mathbf{d=8}$ & $\mathbf{d=1}$ & $\mathbf{d=2}$ & $\mathbf{d=4}$ \\
\hline
\multicolumn{1}{|c|}{\textbf{\DirectCacheFMAName}                   } &     448.79 &     847.67 &     814.03 &     414.77 &     652.56 &     668.08 \\
\multicolumn{1}{|c|}{\textbf{\DirectFMAName}                        } &     440.35 &     823.18 &     732.21 &     421.87 &     695.21 &     674.14 \\
\multicolumn{1}{|c|}{\textbf{\DirectGapFMAName}                       } &     267.34 &     708.09 &     685.85 &     257.32 &     453.12 &     583.28 \\
\multicolumn{1}{|c|}{\textbf{\DirectCacheName}                      } &     412.61 &     790.21 &     780.62 &     406.19 &     590.19 &     625.65 \\
\multicolumn{1}{|c|}{\textbf{\DirectName}                           } &     410.28 &     742.86 &     705.74 &     422.84 &     637.41 &     629.33 \\
\multicolumn{1}{|c|}{\textbf{\DirectGapName}                          } &     246.33 &     646.10 &     661.70 &     246.63 &     426.75 &     549.45 \\
\multicolumn{1}{|c|}{\textbf{\NonaryName}                           } &     247.05 &        --- &        --- &        --- &        --- &        --- \\
\multicolumn{1}{|c|}{\textbf{\PentaryName}                          } &     262.06 &        --- &        --- &     254.42 &        --- &        --- \\
\multicolumn{1}{|c|}{\textbf{\TernaryName}                          } &        --- &        --- &        --- &     159.89 &        --- &        --- \\
\multicolumn{1}{|c|}{\textbf{\EytzingerName}                        } &     174.00 &     366.10 &     401.22 &     174.76 &     253.34 &     284.29 \\
\multicolumn{1}{|c|}{\textbf{\BitSetName}                           } &     202.63 &     374.59 &     399.67 &     208.22 &     261.19 &     219.84 \\
\multicolumn{1}{|c|}{\textbf{\ClassicOffsetName}                    } &     183.88 &     369.72 &     399.58 &     186.30 &     250.59 &     220.26 \\
\multicolumn{1}{|c|}{\textbf{\MorinOffsetName}                      } &     145.07 &        --- &        --- &     146.76 &        --- &        --- \\
\multicolumn{1}{|c|}{\textbf{\BitSetNoPadName}                      } &     160.19 &     364.79 &     392.95 &     164.73 &     244.75 &     215.91 \\
\multicolumn{1}{|c|}{\textbf{\ClassicModName}                       } &     135.92 &     215.61 &     248.52 &     135.98 &     134.80 &     136.88 \\
\multicolumn{1}{|c|}{\textbf{\MorinBranchyName}                     } &      59.42 &        --- &        --- &      60.05 &        --- &        --- \\
\multicolumn{1}{|c|}{\textbf{\ClassicName}                          } &      49.99 &        --- &        --- &      50.72 &        --- &        --- \\
\multicolumn{1}{|c|}{\textbf{\LowerBoundName}                       } &      59.31 &        --- &        --- &      59.81 &        --- &        --- \\
\multicolumn{1}{|c|}{\textbf{\MKLName}                              } &      33.51 &     195.60 &     195.46 &      32.17 &     216.80 &     216.54 \\
\hline
\end{tabular}
\caption{Throughput in millions of searches per second with vector $X$ of size 15}
\label{tab:results0}
\end{table}

\begin{table}[ht]
\centering
\footnotesize

\begin{tabular}{l | c c c | c c c |}
\cline{2-7}
              & \multicolumn{3}{c|}{\textbf{Single}} & \multicolumn{3}{c|}{\textbf{Double}} \\
\cline{2-7}
              & \textbf{Scalar} & \textbf{SSE-4} & \textbf{AVX-2} & \textbf{Scalar} & \textbf{SSE-4} & \textbf{AVX-2} \\
              & $\mathbf{d=1}$ & $\mathbf{d=4}$ & $\mathbf{d=8}$ & $\mathbf{d=1}$ & $\mathbf{d=2}$ & $\mathbf{d=4}$ \\
\hline
\multicolumn{1}{|c|}{\textbf{\DirectCacheFMAName}                   } &     443.32 &     855.97 &     847.05 &     418.69 &     650.46 &     669.65 \\
\multicolumn{1}{|c|}{\textbf{\DirectFMAName}                        } &     437.25 &     825.10 &     746.88 &     421.13 &     684.48 &     683.87 \\
\multicolumn{1}{|c|}{\textbf{\DirectGapFMAName}                       } &     271.68 &     729.22 &     731.69 &     255.48 &     456.37 &     586.83 \\
\multicolumn{1}{|c|}{\textbf{\DirectCacheName}                      } &     422.44 &     811.07 &     790.63 &     409.17 &     583.29 &     633.31 \\
\multicolumn{1}{|c|}{\textbf{\DirectName}                           } &     416.98 &     771.24 &     718.28 &     420.32 &     637.30 &     621.33 \\
\multicolumn{1}{|c|}{\textbf{\DirectGapName}                          } &     252.03 &     657.39 &     678.46 &     247.11 &     425.14 &     549.04 \\
\multicolumn{1}{|c|}{\textbf{\NonaryName}                           } &     154.32 &        --- &        --- &        --- &        --- &        --- \\
\multicolumn{1}{|c|}{\textbf{\PentaryName}                          } &     109.78 &        --- &        --- &     104.57 &        --- &        --- \\
\multicolumn{1}{|c|}{\textbf{\TernaryName}                          } &        --- &        --- &        --- &      65.63 &        --- &        --- \\
\multicolumn{1}{|c|}{\textbf{\EytzingerName}                        } &      67.86 &     121.27 &     149.79 &      67.54 &      88.13 &      95.89 \\
\multicolumn{1}{|c|}{\textbf{\BitSetName}                           } &      82.52 &     120.27 &     151.18 &      81.38 &      89.35 &      83.09 \\
\multicolumn{1}{|c|}{\textbf{\ClassicOffsetName}                    } &      73.46 &     120.52 &     150.94 &      73.45 &      88.89 &      82.92 \\
\multicolumn{1}{|c|}{\textbf{\MorinOffsetName}                      } &      61.12 &        --- &        --- &      61.25 &        --- &        --- \\
\multicolumn{1}{|c|}{\textbf{\BitSetNoPadName}                      } &      68.62 &     120.40 &     149.82 &      68.95 &      86.26 &      82.16 \\
\multicolumn{1}{|c|}{\textbf{\ClassicModName}                       } &      53.32 &      86.57 &     115.73 &      53.37 &      57.57 &      63.44 \\
\multicolumn{1}{|c|}{\textbf{\MorinBranchyName}                     } &      22.07 &        --- &        --- &      21.99 &        --- &        --- \\
\multicolumn{1}{|c|}{\textbf{\ClassicName}                          } &      20.98 &        --- &        --- &      21.15 &        --- &        --- \\
\multicolumn{1}{|c|}{\textbf{\LowerBoundName}                       } &      20.77 &        --- &        --- &      20.82 &        --- &        --- \\
\multicolumn{1}{|c|}{\textbf{\MKLName}                              } &      25.15 &     135.67 &     136.72 &      22.98 &     126.52 &     125.51 \\
\hline
\end{tabular}
\caption{Throughput in millions of searches per second with vector $X$ of size 255}
\label{tab:results1}
\end{table}

\begin{table}[ht]
\centering
\footnotesize

\begin{tabular}{l | c c c | c c c |}
\cline{2-7}
              & \multicolumn{3}{c|}{\textbf{Single}} & \multicolumn{3}{c|}{\textbf{Double}} \\
\cline{2-7}
              & \textbf{Scalar} & \textbf{SSE-4} & \textbf{AVX-2} & \textbf{Scalar} & \textbf{SSE-4} & \textbf{AVX-2} \\
              & $\mathbf{d=1}$ & $\mathbf{d=4}$ & $\mathbf{d=8}$ & $\mathbf{d=1}$ & $\mathbf{d=2}$ & $\mathbf{d=4}$ \\
\hline
\multicolumn{1}{|c|}{\textbf{\DirectCacheFMAName}                   } &     380.30 &     782.41 &     637.10 &     389.57 &     561.19 &     591.66 \\
\multicolumn{1}{|c|}{\textbf{\DirectFMAName}                        } &     392.71 &     683.51 &     597.80 &     363.94 &     556.46 &     542.64 \\
\multicolumn{1}{|c|}{\textbf{\DirectGapFMAName}                       } &     230.93 &     628.40 &     625.06 &     216.89 &     385.00 &     501.90 \\
\multicolumn{1}{|c|}{\textbf{\DirectCacheName}                      } &     359.03 &     738.26 &     636.12 &     369.11 &     523.45 &     556.97 \\
\multicolumn{1}{|c|}{\textbf{\DirectName}                           } &     372.00 &     624.38 &     573.23 &     382.13 &     528.41 &     532.56 \\
\multicolumn{1}{|c|}{\textbf{\DirectGapName}                          } &     223.63 &     583.07 &     596.67 &     208.23 &     362.66 &     481.57 \\
\multicolumn{1}{|c|}{\textbf{\NonaryName}                           } &      98.61 &        --- &        --- &        --- &        --- &        --- \\
\multicolumn{1}{|c|}{\textbf{\PentaryName}                          } &      56.91 &        --- &        --- &      55.00 &        --- &        --- \\
\multicolumn{1}{|c|}{\textbf{\TernaryName}                          } &        --- &        --- &        --- &      44.79 &        --- &        --- \\
\multicolumn{1}{|c|}{\textbf{\EytzingerName}                        } &      36.90 &      66.26 &      87.03 &      37.57 &      45.09 &      53.85 \\
\multicolumn{1}{|c|}{\textbf{\BitSetName}                           } &      44.02 &      66.42 &      88.66 &      43.56 &      46.12 &      48.31 \\
\multicolumn{1}{|c|}{\textbf{\ClassicOffsetName}                    } &      41.62 &      66.47 &      88.17 &      40.79 &      46.48 &      48.73 \\
\multicolumn{1}{|c|}{\textbf{\MorinOffsetName}                      } &      35.38 &        --- &        --- &      34.75 &        --- &        --- \\
\multicolumn{1}{|c|}{\textbf{\BitSetNoPadName}                      } &      39.00 &      66.52 &      88.10 &      39.33 &      44.39 &      48.07 \\
\multicolumn{1}{|c|}{\textbf{\ClassicModName}                       } &      30.52 &      51.26 &      71.73 &      30.64 &      32.24 &      39.02 \\
\multicolumn{1}{|c|}{\textbf{\MorinBranchyName}                     } &      13.59 &        --- &        --- &      13.86 &        --- &        --- \\
\multicolumn{1}{|c|}{\textbf{\ClassicName}                          } &      13.20 &        --- &        --- &      13.54 &        --- &        --- \\
\multicolumn{1}{|c|}{\textbf{\LowerBoundName}                       } &      12.88 &        --- &        --- &      13.09 &        --- &        --- \\
\multicolumn{1}{|c|}{\textbf{\MKLName}                              } &      20.71 &      63.93 &      63.88 &      18.68 &      65.16 &      65.02 \\
\hline
\end{tabular}
\caption{Throughput in millions of searches per second with vector $X$ of size 4095}
\label{tab:results2}
\end{table}

\begin{table}[ht]
\centering
\footnotesize

\begin{tabular}{l | c c c | c c c |}
\cline{2-7}
              & \multicolumn{3}{c|}{\textbf{Single}} & \multicolumn{3}{c|}{\textbf{Double}} \\
\cline{2-7}
              & \textbf{Scalar} & \textbf{SSE-4} & \textbf{AVX-2} & \textbf{Scalar} & \textbf{SSE-4} & \textbf{AVX-2} \\
              & $\mathbf{d=1}$ & $\mathbf{d=4}$ & $\mathbf{d=8}$ & $\mathbf{d=1}$ & $\mathbf{d=2}$ & $\mathbf{d=4}$ \\
\hline
\multicolumn{1}{|c|}{\textbf{\DirectCacheFMAName}                   } &     329.84 &     648.44 &     575.11 &     307.91 &     501.34 &     534.68 \\
\multicolumn{1}{|c|}{\textbf{\DirectFMAName}                        } &     214.62 &     416.77 &     417.74 &     193.22 &     328.62 &     340.12 \\
\multicolumn{1}{|c|}{\textbf{\DirectGapFMAName}                       } &     153.51 &     395.74 &     422.29 &     133.26 &     239.98 &     310.66 \\
\multicolumn{1}{|c|}{\textbf{\DirectCacheName}                      } &     318.65 &     631.74 &     555.77 &     315.58 &     451.57 &     491.66 \\
\multicolumn{1}{|c|}{\textbf{\DirectName}                           } &     226.59 &     398.87 &     408.27 &     205.92 &     320.68 &     326.78 \\
\multicolumn{1}{|c|}{\textbf{\DirectGapName}                          } &     147.70 &     377.56 &     407.99 &     131.61 &     225.11 &     308.79 \\
\multicolumn{1}{|c|}{\textbf{\NonaryName}                           } &      53.79 &        --- &        --- &        --- &        --- &        --- \\
\multicolumn{1}{|c|}{\textbf{\PentaryName}                          } &      44.95 &        --- &        --- &      39.48 &        --- &        --- \\
\multicolumn{1}{|c|}{\textbf{\TernaryName}                          } &        --- &        --- &        --- &      23.73 &        --- &        --- \\
\multicolumn{1}{|c|}{\textbf{\EytzingerName}                        } &      22.97 &      44.45 &      53.46 &      18.33 &      26.88 &      30.00 \\
\multicolumn{1}{|c|}{\textbf{\BitSetName}                           } &      22.34 &      31.06 &      42.54 &      12.18 &      11.14 &      15.04 \\
\multicolumn{1}{|c|}{\textbf{\ClassicOffsetName}                    } &      20.46 &      31.09 &      42.44 &      11.55 &      11.08 &      14.97 \\
\multicolumn{1}{|c|}{\textbf{\MorinOffsetName}                      } &      18.01 &        --- &        --- &      10.13 &        --- &        --- \\
\multicolumn{1}{|c|}{\textbf{\BitSetNoPadName}                      } &      19.68 &      31.03 &      42.33 &      11.15 &      10.98 &      14.98 \\
\multicolumn{1}{|c|}{\textbf{\ClassicModName}                       } &      14.54 &      26.26 &      37.57 &       8.50 &       9.23 &      13.23 \\
\multicolumn{1}{|c|}{\textbf{\MorinBranchyName}                     } &       9.37 &        --- &        --- &       6.86 &        --- &        --- \\
\multicolumn{1}{|c|}{\textbf{\ClassicName}                          } &       9.04 &        --- &        --- &       6.70 &        --- &        --- \\
\multicolumn{1}{|c|}{\textbf{\LowerBoundName}                       } &       8.85 &        --- &        --- &       6.58 &        --- &        --- \\
\multicolumn{1}{|c|}{\textbf{\MKLName}                              } &      16.41 &      31.69 &      31.64 &      13.12 &      25.51 &      25.52 \\
\hline
\end{tabular}
\caption{Throughput in millions of searches per second with vector $X$ of size 65535}
\label{tab:results3}
\end{table}

\begin{table}[ht]
\centering
\footnotesize

\begin{tabular}{l | c c c | c c c |}
\cline{2-7}
              & \multicolumn{3}{c|}{\textbf{Single}} & \multicolumn{3}{c|}{\textbf{Double}} \\
\cline{2-7}
              & \textbf{Scalar} & \textbf{SSE-4} & \textbf{AVX-2} & \textbf{Scalar} & \textbf{SSE-4} & \textbf{AVX-2} \\
              & $\mathbf{d=1}$ & $\mathbf{d=4}$ & $\mathbf{d=8}$ & $\mathbf{d=1}$ & $\mathbf{d=2}$ & $\mathbf{d=4}$ \\
\hline
\multicolumn{1}{|c|}{\textbf{\DirectCacheFMAName}                   } &     139.13 &     135.05 &     135.20 &     109.20 &     112.05 &     110.22 \\
\multicolumn{1}{|c|}{\textbf{\DirectFMAName}                        } &      84.87 &      82.34 &      82.00 &      71.44 &      68.96 &      69.30 \\
\multicolumn{1}{|c|}{\textbf{\DirectGapFMAName}                       } &      88.05 &     100.18 &      99.50 &      75.49 &      81.65 &      80.34 \\
\multicolumn{1}{|c|}{\textbf{\DirectCacheName}                      } &     140.52 &     133.05 &     134.88 &     111.20 &     109.93 &     111.80 \\
\multicolumn{1}{|c|}{\textbf{\DirectName}                           } &      84.17 &      82.48 &      82.16 &      70.00 &      68.63 &      69.06 \\
\multicolumn{1}{|c|}{\textbf{\DirectGapName}                          } &      87.66 &     100.00 &      99.47 &      75.70 &      81.27 &      80.21 \\
\multicolumn{1}{|c|}{\textbf{\NonaryName}                           } &      32.99 &        --- &        --- &        --- &        --- &        --- \\
\multicolumn{1}{|c|}{\textbf{\PentaryName}                          } &      21.31 &        --- &        --- &      18.04 &        --- &        --- \\
\multicolumn{1}{|c|}{\textbf{\TernaryName}                          } &        --- &        --- &        --- &      12.58 &        --- &        --- \\
\multicolumn{1}{|c|}{\textbf{\EytzingerName}                        } &      10.10 &      18.78 &      24.79 &       8.74 &      11.37 &      13.38 \\
\multicolumn{1}{|c|}{\textbf{\BitSetName}                           } &       5.88 &      11.04 &      17.75 &       5.23 &       5.74 &       8.43 \\
\multicolumn{1}{|c|}{\textbf{\ClassicOffsetName}                    } &       5.87 &      11.04 &      17.76 &       5.25 &       5.71 &       8.41 \\
\multicolumn{1}{|c|}{\textbf{\MorinOffsetName}                      } &       4.76 &        --- &        --- &       4.29 &        --- &        --- \\
\multicolumn{1}{|c|}{\textbf{\BitSetNoPadName}                      } &       5.18 &      11.03 &      17.84 &       4.69 &       5.68 &       8.41 \\
\multicolumn{1}{|c|}{\textbf{\ClassicModName}                       } &       4.16 &       9.96 &      16.44 &       3.81 &       4.95 &       7.70 \\
\multicolumn{1}{|c|}{\textbf{\MorinBranchyName}                     } &       4.38 &        --- &        --- &       3.82 &        --- &        --- \\
\multicolumn{1}{|c|}{\textbf{\ClassicName}                          } &       4.31 &        --- &        --- &       3.76 &        --- &        --- \\
\multicolumn{1}{|c|}{\textbf{\LowerBoundName}                       } &       4.23 &        --- &        --- &       3.71 &        --- &        --- \\
\multicolumn{1}{|c|}{\textbf{\MKLName}                              } &       8.20 &      11.59 &      11.61 &       7.26 &      10.40 &      10.37 \\
\hline
\end{tabular}
\caption{Throughput in millions of searches per second with vector $X$ of size 1048575}
\label{tab:results4}
\end{table}

\end{document}